\documentclass[aps,reprint,twocolumn,pre,showpacs,superscriptaddress,floatfix]{revtex4-1}
\usepackage{amssymb,amsmath,amsfonts} 
\usepackage{epsfig,graphicx}
\begin{document}

\title{Entropy, chaos and excited-state quantum phase transitions in the Dicke model}

\author{C. M. L\'obez} \affiliation{Departamento de F\'{\i}sica
  Aplicada I and GISC, Universidad Complutense de Madrid,
  Av. Complutense s/n, 28040 Madrid, Spain} \author{A.
  Rela\~{n}o} \affiliation{Departamento de F\'{\i}sica Aplicada I and
  GISC, Universidad Complutense de Madrid, Av. Complutense s/n, 28040
  Madrid, Spain}

\begin{abstract}
  We study non-equilibrium processes in an isolated quantum system
  ---the Dicke model--- focusing on the role played by the transition
  from integrability to chaos and the presence of excited-state
  quantum phase transitions. We show that both diagonal and
  entanglement entropies are abruptly increased by the onset of
  chaos. Also, this increase ends in both cases just after the system
  crosses the critical energy of the excited-state quantum phase
  transition. The link between entropy production, the development of
  chaos and the excited-state quantum phase transition is more clear
  for the entanglement entropy. On the contrary, the heat dissipated
  by the process is affected neither by the onset of chaos, nor by the
  excited-state quantum phase transition.
\end{abstract}


\maketitle

\section{Introduction}
 
The relation between statistical mechanics and the underlying
microscopic dynamics is still object of discussion. During the last
couple of years, the development of experimental techniques allowing
to manipulate isolated quantum systems with high degree of accuracy
and control has spurred on the fundamental research both
experimentally and theoretically \cite{Bloch:08,Polkovnikov:11b}. In
particular, the consequences of performing a quench ---a sudden change
in the external parameters of the Hamiltonian--- have been largely
explored, focusing on different fundamental open questions: the
process of thermalization \cite{Deutsch:91, Srednicki:94,Rigol:08},
the irreversibility arising from non-equilibrium dynamics
\cite{QuantumThermodynamics,Esposito:10,Allahverdyan:05,Puebla:15}, or
the consequences of crossing critical points
\cite{Polkovnikov:05,Zurek:05,Grandi:09}, just to cite a few. This
paper follows the same line of research. We rely on the paradigmatic
Dicke model \cite{Dicke:54} to study how entropy and entanglement are
produced by a quench, when the dynamics is affected by the onset of
chaos and the presence of excited-state quantum phase transitions
(ESQPTs). Our main conclusion is that all these phenomena are closely
connected. Entropy production and entanglement are enhanced by the
emergence of chaos and, though in a less clear way, by the presence of
an ESQPT. On the other hand, the energy dissipated or {\em lost} by
the quench is affected by none of these phenomena.

In the macroscopic realm, irreversibility, the corresponding increase
of entropy, and a number of statements about the work lost as a
consequence of a process are equivalent \cite{termo}. However, the
situation is totally different in a (small) isolated quantum system
following unitary time evolution. The energy dissipated or lost by a
non-equilibrium process can be univocally calculated (see below for
details). But the von Neumann entropy of the system, $S = - \text{Tr}
\rho \log \rho$, is always constant; it does not provide a microscopic
basis for the Second Law. In other words, any time-dependent process
$H(t)$ in any isolated quantum system keeps all the information about
the initial state, and hence no entropy is produced as a consequence
of it.

A number ways have been recently explored to solve this
problem. One consistent alternative relies on the diagonal entropy
\cite{Polkovnikov:11}. If the time-dependent wavefunction of a system
following a non-equilibrium process is written $\left| \psi(t) \right>
= \sum_n C_n (t) \left| E_n (t) \right>$, where $\left| E_n (t)
\right>$ are the instantaneous eigenstates of the system, $H(t) \left|
  E_n (t) \right> = E_n (t) \left| E_n (t) \right>$, the diagonal
entropy is defined $S_d = - \sum_n \left| C_n (t) \right|^2 \log
\left| C_n (t) \right|^2$. It can be understood as a natural
consequence of the equilibration process that any isolated quantum
system experiments after a change in its external parameters ---the
process after which the state $\left| \psi(t) \right>$ {\em relaxes}
to an equilibrium mixed state $\rho_{\text{eq}} = \lim_{T \rightarrow
  \infty} \int_0^T dt \, \left| \psi(t) \right> \left< \psi(t)
\right|$, around which it fluctuates during the majority of the
time \cite{Reimann:12}. As a consequence of the long-time average, the
non-diagonal elements of $\rho_{\text{eq}}$ vanish, and hence
$\rho_{\text{eq}}$ reduces to a diagonal mixed state,
$\rho_{\text{eq}} = \sum_n \left| C_n \right|^2 \left| E_n \right>
\left< E_n \right|$. So, the diagonal entropy can be understood as the
von Neumann entropy of this effective equilibrium state, resulting
from dephasing. This entropy shares a number of important properties
with the thermodynamic entropy. For example, in a time-dependent
process $H(t)$, $S_d$ only remains constant in the adiabatic limit;
the adiabatic theorem forbids transitions between the different
instantaneous eigenstates, so the coefficients $C_n(t)$
only acquire phases during the whole process. Furthermore, it is
expected to provide a good description of any experiment performed
over the system; if the system thermalizes, the effective equilibrated
state, $\rho_{\text{eq}}$, is expected to reproduce the expected
values of physical observables, $\left< {\mathcal O} \right> \sim
\text{Tr} \left[ {\mathcal O \rho_{\text{eq}}} \right]$.
Unfortunately, it also suffers from some important limitations. For
example, it changes immediately after a sudden quench $H(\lambda_i)
\longrightarrow H(\lambda_f)$, being $\lambda$ an external control
parameter of the Hamiltonian ---it changes as soon as the quench is
completed, not after the corresponding relaxation time. Hence, if the
system experiments a very fast cycle $H(\lambda_i) \longrightarrow
H(\lambda_f) \longrightarrow H(\lambda_i)$, with a characteristic time
much shorter that the relaxation time, this entropy first increases
after the forward part of the cycle, and then decreases again after
the backward. The reason is that the diagonal entropy $S_d$ is equal
to de von Neumann entropy of the effective equilibration state
$\rho_{\text{eq}}$, so it implicitly assumes that the relaxation has
been completed.

Another relevant alternative lays on the entanglement between the
system under study and its environment \cite{review}. The most
distinctive feature of entanglement, the fact that none of the parts
of an entangled system exists as an individual subsystem, has been
resorted to link the Second Law and the unitary quantum evolution
\cite{Esposito:10,Popescu:06,Khlebnikov:14,Kaufman:16}. Let's consider
that we deal with a global pure state, describing what we call the
{\em universe}, that evolves unitarily under a certain global
Hamiltonian $H$. All the information about this global state is
conserved by the time evolution, and thus it has no entropy ---despite
it remains close to the effective relaxed state $\rho_{\text{eq}}$
during the majority of the time. However, being this true, all the
possible experiments we can perform over this global state are
restricted to a small part of it ---what we call the {\em system}. And
due to the entanglement between the different parts of the global pure
state, this fact entails that we lack a relevant piece of information
regarding the possible outcomes of our experiments, information that
is encoded in the (inaccessible) correlations between the system and
the rest of the universe ---what we call the {\em
  environment}. Therefore, from the results of all our possible
experiments we infer that our system is in a mixed state, follows a
non-unitary time-evolution and has a (thermodynamic) entropy
growing in time. In other words, quantum correlations between the
system and its environment entail an objective and unavoidable loss of
information about the state of the system, and thus they can be the
ultimate responsible of the Second Law. Indeed, these quantum
correlations have been shown to be under the emergence of canonical
states in small parts of large and globally isolated quantum system
\cite{Popescu:06}. Also, entanglement between the different parts of
such systems has been pointed as the mechanism leading to
thermalization \cite{Khlebnikov:14,Kaufman:16}. And the corresponding
entanglement entropy has been shown to split in reversible and
irreversible contributions, the last of which grows as a consequence
of the unitary evolution, provided that the initial condition is a
separable mixed state \cite{Esposito:10}. So, as a promising
alternative for the thermodynamical entropy, we can consider the
reduced density matrix of the system, after tracing out all the
environmental degrees of freedom, $\rho_S (t)= \text{Tr}_E \left|
  \psi(t) \right> \left< \psi(t) \right|$, and its corresponding von
Neumann entropy, $S_{\text{ent}} = - \text{Tr}_S \rho_S (t)$, which
always depends explicitly on time.

The aim of this work is to explore the behavior of both diagonal and
entanglement entropies, in the paradigmatic Dicke model
\cite{Dicke:54}. As we have pointed before, it is reasonable to expect
that both the effective equilibrated state, $\rho_{\text{eq}}$, from
which we obtain the diagonal entropy $S_d$, and the reduced density
matrix, $\rho_S$, giving rise to the entanglement entropy
$S_{\text{ent}}$, provide a good description of the experiments
performed over the system (see below for a precise description about
what we call system and what we call environment), so it is logical to
expect that both entropies account for the irreversibility inherent to
a non-equilibrium process. We have chosen the Dicke model for the following
reasons: {\it i)} it is experimentally accessible, both by means of a
superfluid gas in an optical cavity \cite{Baumann:10}, and by means of
superconducting circuits \cite{Solano:14}; {\it ii)} it is naturally
splitted in two parts: a finite number of atoms (the {\em system}) and
an infinite number of photons (the {\em environment}); and {\it iii)}
its spectrum has a very rich structure, showing both quantum phase
transitions (QPTs) and excited-state quantum phase transitions
(ESQPTs) \cite{Cejnar:06,Caprio:08,Stransky:14}, and a transition from
integrability to chaos as a function of the energy
\cite{Perez-Fernandez:11b,Bastarrachea:14}.

After a stringent numerical study, we obtain the following
conclusions. First, the energy dissipated by the quench, $Q$, is
totally oblivious of the presence of critical phenomena and the onset
of chaos. For almost any quench $\lambda_i \rightarrow \lambda_f$ (see
below for details), the dissipated energy is proportional to the
quench size $\Delta \lambda \equiv \lambda_f - \lambda_i$, $Q \sim 2
\left( \Delta \lambda \right)^2$, no matters whether the final
Hamiltonian has a coupling constant $\lambda$ above or below the
critical, $\lambda_c$, and whether the final energy $E$ is below or
above the corresponding to the onset of chaos and the critical energy
$E_c$ of the ESQPT. This means that the structure of the spectrum is
irrelevant from a practical point of view; the energy that a machine
operating between $\lambda_i$ and $\lambda_f$ will lose as a
consequence of irreversibility only depends on the size of the quench,
neither on the chaotic or regular nature of the dynamics, nor on the
presence of an ESQPT. On the contrary, both diagonal and entanglement
entropies are abruptly increased in the surroundings of $E_c$, a
region in which the system transits from almost integrable to fully
chaotic dynamics. A possible explanation is that a number of
approximately conserved quantities holding below the onset of chaos
restrict the dynamics of the system, making the entropy increase less
than expected in this region. It is also worth to stress that diagonal
and entanglement entropies do not behave in the same way. A tiny
quench is enough to increase the first one; non-adiabatic transitions
resulting from the quench spread the final energy distribution over
many different eigenstates and give rise to a large value of the
diagonal entropy, no matters the approximated conservation rules
holding below the onset of chaos. On the contrary, entanglement is
almost zero until the system approaches the critical energy of the
ESQPT and enters in the region in which the chaotic behavior
starts. So, the behavior of these entropies is determined by different
physical mechanisms. Diagonal entropy responds to any non-adiabatic
transition between energy levels, providing a good estimate of
irreversibility after the system is equilibrated in its final
state. Entanglement entropy is more sensitive to structural changes in
the eigenstates of the final Hamiltonian, and in particular to the
mechanisms expected to lead the system to a final thermal
state. Hence, from a thermodynamic point of view, diagonal entropy is
more adequate because is the only which captures the consequences of
all the non-adiabatic transitions between energy levels; however, as
we have pointed above, it is not useful to derive a time-dependent
entropy $S(t)$, supporting the Second Law. On the other hand,
entanglement entropy does increase in time, at least for quenches
large enough; but it does not account for all the irreversible
changes.

The rest of the paper is organized as follows. In section \ref{model} we
present the Dicke model, discuss its main features, and describe the
process from which we obtain the results. In section \ref{results} we
show the main results of the paper. Finally, in section \ref{conclusions}
we summarize our main conclusions.

\section{Model and protocol}
\label{model}

The Dicke model \cite{Dicke:54} describes the interaction of $N$
two-level atoms of splitting $\omega_0$ with a single bosonic mode of
frequency $\omega$, by means of a coupling parameter $\lambda$,
\begin{equation}
H = \omega_0 J_z + \omega a^{\dagger} a + \frac{2 \lambda}{\sqrt{N}} J_x \left( a^{\dagger} + a \right).
\label{eq:Dicke}
\end{equation}
In this representation, $a^{\dagger}$ and $a$ are the creation and
annihilation operators of photons, and $J = (J_x ,J_y ,J_z )$ is the
total angular momentum. This Hamiltonian has two main conserved
quantities. First, $J^2$, which divides the Hamiltonian in diagonal
blocks; in this paper we restrict ourselves to $J=N/2$, which is
enough to deal with the recent experimental realizations
\cite{Baumann:10,Solano:14}. Second, due to the invariance of $H$
under $J_x \rightarrow - J_x$ and $a \rightarrow - a$, $\Pi = \exp
\left( i \pi \left[ J + J_z + a^{\dagger} a \right] \right)$ is also a
conserved quantity. As this is a discrete symmetry, $\Pi$ has only two
different eigenvalues, $\Pi \left|E_i, \pm \right> = \pm \left|E_i,
  \pm \right>$, and it is usually called {\em parity}.

The Dicke Hamiltonian undergoes a second-order QPT at $\lambda_c =
\sqrt{\omega \omega_0} /2$, which separates the so-called normal phase
$(\lambda < \lambda_c )$ from the superradiant phase $( \lambda >
\lambda_c )$ \cite{Emary:03,Hepp:73,Wang:73,Carmichael:73}. This
transition has been linked to a singular behavior of the entanglement
between the atoms and the radiation field \cite{Lambert:04}, implying
that an adiabatic evolution from the ground state of the system
results in an abrupt increase of the entanglement at the critical
coupling. It has also been related to the emergence of chaos
\cite{Emary:03}. The region corresponding to the superradiant phase
displays an ESQPT at $E_c = -J \omega_0$. In the thermodynamic limit,
if $\lambda > \lambda_c$ and $E < E_c$, all the energy levels are
doubly degenerate, and therefore parity can be spontaneously broken
\cite{Perez-Fernandez:11,Perez-Fernandez:11b,Puebla:13,Puebla:13b,Brandes:13}. It
has been suggested than the ESQPT induces the development of chaos
\cite{Perez-Fernandez:11b,Relano:14}, but the current results are far
from conclusive \cite{Bastarrachea:14}.

All the results we show in this paper are based on the same
procedure. We consider the ensemble of atoms as the {\it system}, $H_S
= \omega_0 J_z$; the radiation field, as the {\it environment} $H_E =
\omega a^{\dagger} a$; and an interaction between them given by
$H_{int} = 2 \lambda J_x \left( a^{\dagger} + a \right) /
\sqrt{N}$. The unitary time-evolution of any pure state $\left|
  \psi(0) \right>$ is given by $\left| \psi(t) \right> = {\mathcal
  U}(t) \left| \psi(0) \right>$, where ${\mathcal U}(t)$ is the
unitary evolution operator, that can be obtained from
Eq. (\ref{eq:Dicke}). Exact calculations are accessible for current
computational capabilities up to $N \sim 50$, and hence both diagonal
and entanglement entropies can be numerically obtained. For the first
one, we only require to write the time-dependent wavefunction in the
instantaneous eigenbasis, $\left| \psi(t) \right> = \sum_n C_n(t)
\left| E_n(t) \right>$, and taking $S_d = - \sum_n \left| C_n(t)
\right|^2 \log \left| C_n(t) \right|^2$. The entanglement entropy is
obtained by tracing out the photonic degrees of freedom, $\rho_S (t) =
Tr_E \left[ \left| \psi(t) \right> \left< \psi(t) \right| \right]$,
and taking $S_{\text{ent}}(t) = - Tr \left[ \rho_S(t) \log \rho_S(t)
\right]$. It is worth to stress that, due to the correlations
between atoms and photons, $\rho_S (t)$ does not follows an unitary
time evolution, and thus $S_{\text{ent}}(t)$ is explicitly
time-dependent. Furthermore, this entropy has been shown to behave in
the same qualitative way as other entanglement measures, at least for
the ground state of this system \cite{Lambert:04}.

Our main objective is to explore the consequences of a highly
non-equilibrium process, by suddenly changing the external parameter
from $\lambda_i$ to $\lambda_f$. This procedure is very easy to
simulate, because the corresponding Hamiltonian does not depend
explicitly on time; the exact time evolution comes directly from the
final value of the coupling constant $\lambda_f$, $\left| \psi(t)
\right> = e^{i H \left(\lambda_f \right) t} \left| \psi(0) \right>$,
  where the initial condition $\left| \psi(0) \right>$ is the ground
  state of the system at the initial value of the coupling constant
  $\lambda_i$. We consider $\omega=\omega_0=\hbar=1$ throughout all
  this paper. Also we always take $\lambda_i > \lambda_f > \lambda_c$,
  in order to explore the complete phase diagram of the Dicke model
  (see \cite{Puebla:13} for details). 

  Another advantage of this procedure is that the ground state is very
  well described by a separable coherent state, $\left| \mu, \nu
  \right> = \left| \mu \right> \otimes \left| \nu \right>$
  \cite{Zhang:90}, where
\begin{eqnarray}
\left| \mu \right> &=& \left( 1 + \mu^2)^{-J} \exp( \mu J_+ \right) \left| J, -J \right>, \\
\left| \nu \right> &=& \exp \left( \nu^2/2 \right) \exp \left( \nu a^{\dagger} \right) \left| 0 \right>,
\end{eqnarray}
correspond to the atomic and the bosonic parts of the state,
respectively. The precise values of both parameters for a certain
coupling constant $\lambda_i$ are obtained minimizing the energy
surface $\left< \mu, \nu \right| H(\lambda_i) \left| \mu, \nu
\right>$,
\begin{eqnarray}
\mu &=& \pm \sqrt{ \frac{\lambda_i^2 - \lambda_c^2}{\lambda_i^2 + \lambda_c^2}}, \nonumber \\
\nu &=& \mp \frac{\sqrt{2J}}{\omega} \frac{\sqrt{\lambda_i^4 - \lambda_c^4}}{\lambda},
\label{eq:munu}
\end{eqnarray}
considering that we are always in the superradiant phase $\lambda_i >
\lambda_c$. Furthermore, as in this phase $(\lambda > \lambda_c)$ the
ground state is always degenerate, any linear combination of the
previous coherent states is also a valid ground state,
\begin{equation}
\left| \mu, \nu \right> = \frac{ \alpha \left| \mu_+ \right> \otimes \left| \nu_- \right> \pm \beta \left| \mu_- \right> \otimes \left| \nu_+ \right>}{\sqrt{\left| \alpha \right|^2 + \left| \beta \right|^2}},
\label{eq:general}
\end{equation}
where the subindex $+$ ($-$) indicate the positive (negative) values
in (\ref{eq:munu}). Throughout all this work we restrict ourselves to
the case $\alpha=1$, $\beta=0$. As this is a product state, all the
entanglement observed in the final equilibrium state is produced by
the non-equilibrium process. Similar qualitative results are obtained
for other values for $\alpha$ and $\beta$ (not shown). The only
difference is that the generic state in (\ref{eq:general}) is yet
entangled, and thus its initial value for $S_{\text{ent}}$ is not
zero; a straightforward calculation shows that
$S_{\text{ent}}(0)=\log\left( \left| \alpha \right|^2 + \left| \beta
  \right|^2 \right) - \frac{1}{\left| \alpha \right|^2 + \left| \beta
  \right|^2} \left( \left| \alpha \right|^2 \log \left| \alpha
  \right|^2 + \left| \beta \right|^2 \log \left| \beta \right|^2
\right)$. Initial states with $\alpha = 1 / \sqrt{2}$ and $\beta = \pm
1 / \sqrt{2}$ are particularly interesting because have well-defined
positive or negative parity; in both cases $S_{\text{ent}}(0) = \log
2$. Note that for all these situations, $S_d(0)=0$, since we always
start the time evolution from the ground state.

A thermodynamic description of the process is done in the following
terms. As the complete system remains always thermally isolated, we
consider the external parameter $\lambda$, the internal energy $E$ and
the entropy $S$, as the thermodynamic variables. Any quench with
$\lambda_i > \lambda_f$ requires a certain amount of mechanical work
(see \cite{Puebla:13} for details), which entails an increase of the
internal energy $\Delta E = \left<\psi(0) \right| H(\lambda_f) \left|
  \psi(0) \right> - \left<\psi(0) \right| H(\lambda_i) \left| \psi(0)
\right>$. To determine which part of this energy is lost as a
consequence of the irreversibility of the process, we can consider
that any (differential) energy change can be splitted in two terms, $d
E = \sum_n P_n d E_n + E_n d P_n$, where $E_n$ are the eigenvalues of
the system, and $P_n$ the probabilities of obtaining each one in an
energy measurement \cite{Quan:05,Polkovnikov:08}. The first term in
the sum is identified as the (reversible) work done by the quench, $W
= \sum_n P_n d E_n$, and the second one as the heat dissipated by the
quench, $Q = \sum_n E_n d P_n$. As it is stated by the adiabatic
theorem, if the process is completed slowly enough there are no
transitions between different energy levels, implying $Q=0$ and
$\Delta E = W$. On the contrary, irreversible protocols entail
non-adiabatic transitions between energy levels, and therefore
$Q>0$. Hence, $Q$ can be understood as the mechanical work lost by the
protocol. It comes from a simple (and reversible) mechanical change
$\lambda_i \rightarrow \lambda_f$, but entails a large number of
microscopic transitions between energy levels, that cannot be
retrieved without a precise microscopic knowledge of the wavefunction.

The aim of this paper is to study how diagonal and
entanglement entropies increase as a consequence of a quench, and
hence if they provide good microscopic basis for the thermodynamic
entropy. 

\section{Numerical results}
\label{results}

First, we study the behavior of the system with $N=30$ atoms, and
different values of the final coupling constant. In
Fig. \ref{fig:heat} we show the results for the dissipated
energy. Symbols show the numerical results for $\lambda_f=0.9$, $1.2$,
$1.5$, $2.0$, $2.5$, and $3.0$ (see caption of Fig. \ref{fig:heat}
for details). The most representative fact is that the dissipated
energy is the same for all the cases; $Q$ depends just on the size of
the quench, $\Delta \lambda=\lambda_i - \lambda_f$, independently of
the final coupling constant $\lambda_f$. Although not shown, similar
results are obtained for $\lambda_f < \lambda_c$. The dashed line on
the plot represents $Q = 2 \left( \Delta \lambda \right)^2$. This
result comes from the following facts:

\begin{figure}
\begin{center}
\includegraphics[height=0.8\linewidth,angle=-90]{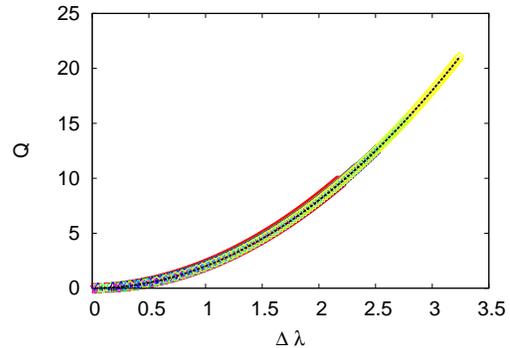}
\end{center}
\caption{Dissipated energy $Q$ for a system with $N=30$ atoms and
  different values for the final coupling constants,
  $\lambda_f$. Results are shown as a function of the size of the
  quench, $\Delta \lambda = \lambda_i - \lambda_f$. Open symbols
  display the numerical results: squares (red online),
  $\lambda_f=0.9$; circles (green online), $\lambda_f=1.2$; upper
  triangles (blue online), $\lambda_f=1.5$; lower triangles (magenta
  online), $\lambda_f=2.0$; diamonds (cyan online), $\lambda_f=2.5$;
  and crosses (yellow online), $\lambda_f=3.0$. Dashed line represents
  $Q = 2 \left( \Delta \lambda \right)^2$.}
\label{fig:heat}
\end{figure}

First, the final energy for a quench $\lambda_i
\rightarrow \lambda_f$ (see \cite{Puebla:13} for details) is,
\begin{equation}
\frac{E \left(\lambda_i, \lambda_f \right)}{J} = \frac{2 \lambda_i \left( \lambda_i - 2 \lambda_f \right)}{\omega} + \frac{\left(2 \lambda_f - 3 \lambda_i \right) \omega \omega_0^2}{8 \lambda_i^3},
\end{equation}
expression valid for $\lambda_i > \lambda_c$. Second, the ground state
energy of the final Hamiltonian is
\begin{equation}
\frac{E_0}{J} = - \omega_0 \frac{\omega \omega_0}{4 \lambda_f^2} - \frac{2}{\omega} \frac{16 \lambda_f^4 - \omega^2 \omega_0^2}{16 \lambda_f^2},
\end{equation}
provided that $\lambda_f > \lambda_c$. From these expressions we obtain,
\begin{equation}
\frac{Q \left(\lambda_i, \lambda_f \right)}{J} = \frac{\left( \lambda_i - \lambda_f \right)^2}{8 \omega} \left[ 16 + \frac{\left( 2 \lambda_f + \lambda_i \right) \omega_0^2 \omega^2}{\lambda_f^2 \lambda_i^3} \right].
\end{equation}
So, considering $\omega=\omega_0=1$ and defining $x \equiv \lambda_i -
\lambda_f$, the dissipated energy reads,
\begin{equation}
\frac{Q \left(x, \lambda_f \right)}{J} = 2 x^2 + \frac{x^2}{8} \frac{3 \lambda_f + x}{\lambda_f^2 \left( \lambda_f+x \right)^3}.
\end{equation}
The main trend of this result is $Q = 2 x^2 = 2 \left( \Delta \lambda
\right)^2$, that is, the corresponding to the dashed line in
Fig. \ref{fig:heat}. The second term in the sum provides a small
correction, negligible for medium and large values of $\Delta
\lambda$.

\begin{figure}
\begin{center}
\includegraphics[height=0.8\linewidth,angle=-90]{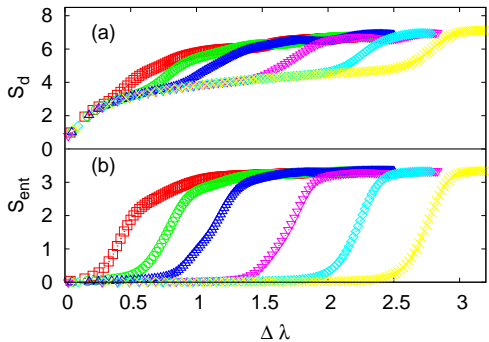}
\end{center}
\caption{Diagonal entropy $S_d$ (panel a) and entanglement entropy
  $S_{\text{ent}}$ (panel b), for a system with $N=30$ atoms and different
  values for the final coupling constants, $\lambda_f$. Results are
  shown as a function of the size of the quench, $\Delta \lambda =
  \lambda_i - \lambda_f$. Open symbols display the numerical results:
  squares (red online), $\lambda_f=0.9$; circles (green online),
  $\lambda_f=1.2$; upper triangles (blue online), $\lambda_f=1.5$;
  lower triangles (magenta online), $\lambda_f=2.0$; diamonds (cyan
  online), $\lambda_f=2.5$; and crosses (yellow online),
  $\lambda_f=3.0$.}
\label{fig:entropies}
\end{figure}

In Fig. \ref{fig:entropies} we show the results for the diagonal
entropy (panel a) and the entanglement entropy (panel b), for the same
cases than in Fig. \ref{fig:heat}, plotted with the same
symbols (see captions for details). Both entropies show significant
differences with respect to the dissipated heat, plotted in
Fig. \ref{fig:heat}. In all the cases, the diagonal entropy $S_d$
fastly increases from $S_d=0$ (limit for quenches small enough to
avoid non-adiabatic transitions), showing that a large number of
energy levels become populated after small or moderate quenches; note
that the entropy generated is the same for all the cases. Then, a
second region appears. Above $\Delta \lambda \sim 0.4$, the curves for
the different cases follow different paths. The entropy generated by
the quench ending at $\lambda_f=0.9$ continues growing with the same
trend, whereas the increase of the others slow down, showing a sort of
plateau. This plateau ends at a different value of $\Delta \lambda$
for each case, giving rise to a second fastly-growing region. The last
part of all the cases consists of another slowly-growing region,
starting at different values of $\Delta \lambda$ for each case: the
larger the $\lambda_f$, the larger $\Delta \lambda$. Summarizing, the
diagonal entropy grows following a complex path which depends on the
final value of the coupling constant, $\lambda_f$.

Results for the entanglement entropy (panel b of
Fig. \ref{fig:entropies}) also show a complex pattern. The first part
of the plots consist of a region in which $S_{\text{ent}} \sim 0$,
that is, a region in which atoms and photons remain unentangled. The
size of this region depends on $\lambda_f$: the larger $\lambda_f$,
the larger the quench $\Delta \lambda$ at which the entanglement
entropy becomes significantly different from zero. Then, a
fastly-growing region starts, very similar to the fastly-growing
region of $S_d$ after the intermediate plateau. In all the cases,
$S_{\text{ent}}$ grows from almost zero to almost its maximum value;
in other words, atoms and photons change from almost pure to almost
maximally entangled states in a narrow band of quench sizes. After
this fastly-growing region, we reach a final regime in which the
entanglement entropy is approximately constant and close to its
maximum possible value, $S_{\text{ent}} = \log 31$.

From these results, we infer the following conclusions. First, the
energy lost by the quench, $Q$, only depends on its size, $\Delta
\lambda$. This entails that the efficiency of a machine operating
between $\lambda_i$ and $\lambda_f$ would depend neither on the final
coupling $\lambda_f$, nor on the final value of the energy $E=\left<
  \psi(0) \right| H \left| \psi(0) \right>$. On the contrary, the
entropy produced by the same process dramatically depends on
$\lambda_f$. As paradigmatic examples, we can compare the cases with
$\lambda_f =0.9$ and $\lambda_f=3.0$. In the first case,
$S_{\text{ent}}$ is roughly at its maximum for $\Delta \lambda \sim
1$, but entailing a small dissipated energy, $Q/J \sim 2$. In the
second one, the same quench size and the same amount of dissipated
energy make atoms and photons remain almost unentangled,
$S_{\text{ent}} \sim 0$. Furthermore, the differences between $S_d$
and $S_{\text{ent}}$ are also significant. For example, the quench
$\Delta \lambda \sim 1$ for the case $\lambda_f = 3.0$, entails $S_d
\sim 3.5$, whereas $S_{\text{ent}} \sim 0$. So, we conjecture that the
entropy growth is not directly related to the dissipated energy, but
to structural changes in the spectrum; and that the entanglement
entropy is more sensitive to this changes. 

In the superradiant phase
($\lambda > \lambda_c$) two main structural changes occur above the
ground state: the onset of chaos and an ESQPT. Recent results suggest
that both phenomena are related (see \cite{Perez-Fernandez:11b} for
the Dicke model, and \cite{Relano:14} for an atom-molecule gas), but
it seems that they do not take place at the very same energy
\cite{Bastarrachea:14}. In any case, the distance between the ground
state and the region in which both the ESQPT and the onset of chaos
take place \cite{Perez-Fernandez:11b, Puebla:13, Bastarrachea:14}
increases with $\lambda_f$, indicating that the size of a quench
$\Delta \lambda$ leading to this region also increases. Hence, it is
logical to look for a connection between the fastly-growing region for
both $S_d$ and $S_{\text{ent}}$, the onset of chaos and the ESQPT.

\subsection{Onset of chaos}

Quantum chaos can be detected by means of several procedures. The
majority of them are related to spectral statistics, that is, to the
statistical properties of the sequence of energy levels
\cite{Gomez:11}. An alternative approach has been used during the last
couple of years to study how the degree of chaos changes with the
excitation energy, a problem which is closely related to the one we
are dealing with \cite{Stransky:09}. It was proposed around 30 years
ago by Peres \cite{Peres:84}. Basically, it consists of drawing the
expected value of a representative observable \cite{nota1} in every
eigenstate, in terms of the corresponding energy. If the system is
fully chaotic, the plot shows no structure. On the contrary, if the
system is integrable the plot is ordered in a regular lattice, due to
the underlying integrals of motion. Such kind of plots are usually
called {\em Peres lattices}.

\begin{figure}
\begin{center}
\includegraphics[height=0.8\linewidth,angle=-90]{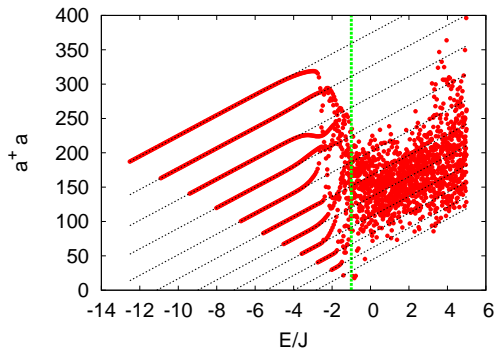}
\end{center}
\caption{Peres Lattice for a system with $\lambda_f=2.5$ and
  $N=30$. Points (red online) represent $\left<E_n \right| a^{\dagger}
  a \left| E_n \right>$ versus the energy $E_n/J$ of the eigenvector
  $\left|E_n \right>$. Vertical dashed line (green online) displays
  the critical energy of the ESQPT, $E_c/J=-1$. Black dashed lines
  show a visual guide to identify the regular pattern in the low-lying
  region of the spectrum, consisting of {\em bands}. These lines are
  obtained by means of linear fit to the points in the corresponding
  band.}
\label{fig:peres}
\end{figure}

In Fig. \ref{fig:peres} we represent a Peres lattice for a case with
$N=30$ and $\lambda_f=2.5$. We have chosen the number of photons,
$a^{\dagger} a$, as a representative observable. Together with the
numerical points, we plot the critical energy of the ESQPT
($E_c/J=-1$), as a vertical dashed (green online) line. Below $E/J
\sim -4$, the lattice is regular. The values $\left<E_n \right|
a^{\dagger} a \left| E_n \right>$ are distributed in bands, each one
following a different linear trend. Dotted lines in
Fig. \ref{fig:peres} represent these linear trends, obtained by
least-squares fits to the corresponding sets of points. Around $E/J
\sim -4$ some bands start to deviate from their linear trends, and
above the critical energy of the ESQPT, $E/J=-1$, all the points
distribute randomly. Thus, we can conclude that the transition from
integrability to chaos happens in this interval. 

Besides the identification of quantum chaos, Peres lattices are also
interesting to study the process of thermalization in isolated quantum
systems. The eigenstate thermalization hypothesis (ETH)
\cite{Rigol:08} states that an isolated quantum system thermalizes if
the expected values of physical observables are approximately the same
in all the states within a small energy window $\Delta E$. This fact
entails that the expected value of any physical observable in a single
eigenstate equals the microcanonical average, and therefore the system
behaves like in thermal equilibrium independently of its actual
probability distribution $P(E_n)$
\cite{Polkovnikov:11b,Deutsch:91,Srednicki:94,Rigol:08}. In our case,
this requirement is not fulfilled for $E/J < -4$; $\left<E_n \right|
a^{\dagger} a \left| E_n \right>$ dramatically changes from band to
band, and hence we cannot expect a thermal behavior in this region. On
the contrary, above the critical energy of the ESQPT $\left<E_n
\right| a^{\dagger} a \left| E_n \right>$ randomly fluctuates around
its average value. The width of the corresponding distribution is
large, but we can expect that its relative size decreases with the
number of atoms $N$, giving rise to thermalization in larger systems.

\begin{figure}
\begin{center}
\includegraphics[height=0.8\linewidth,angle=-90]{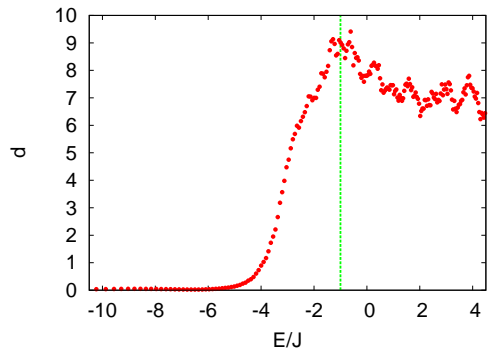}
\end{center}
\caption{Distance $d$ between the points in the Peres lattice of
  Fig. \ref{fig:peres} and their closest linear trend, averaged over
  $\Delta=100$ consecutive points. Vertical dashed line (green online)
  represents the critical energy of the ESQPT, $E/J=-1$.}
\label{fig:chaos}
\end{figure}

A quantitative analysis is presented in Fig. \ref{fig:chaos}, which
has been constructed in the following way. First, we compute the
distance $d$ between every point $\left<E_n \right| a^{\dagger} a
\left| E_n \right>$ in the lattice and its closest linear
trend. Second, we average the distances obtained over a fixed number
of eigenstates, $\Delta$. Finally, we represent the averaged $d$
versus the average energy of the corresponding eigenstates. The
results provide a measure of how ordered is the lattice in the
neighborhood of a given energy $E$. If $d$ is close to zero, the
lattice shows a very regular pattern, and hence the system is almost
integrable at the corresponding energy; if it is clearly different
from zero, the system is far from integrability. Results plotted in
Fig. \ref{fig:chaos} have been obtained with $\Delta=100$. The
distance $d$ remains very close to zero up to energies around $E/J
\sim -4$. Then, it fastly increases up to its maximum, which
remarkably happens around the critical energy of the ESQPT, $E/J \sim
-1$. Finally, the distance $d$ decreases a bit and oscillates around a
certain value. So, we conclude that chaos emerges clearly below $E_c$,
as it was pointed in \cite{Bastarrachea:14}, but the critical energy
still plays a prominent role ---results in Fig. \ref{fig:chaos}
suggest that the system is fully chaotic above the ESQPT. It is worth
to note that the distance $d$ behaves in a similar way than the
entropies shown in Fig. \ref{fig:entropies}, specially
$S_{\text{ent}}$: both $d$ and $S_{\text{ent}}$ remain close to zero
whithin a certain range, and then experiment an abrupt
increase. Though we do not show numerical results, we have checked
that this parallelism also holds for other values of the final coupling
constant $\lambda_f$; for example, if $\lambda_f=0.9$ the transition
to chaos starts at very low excitation energies. Therefore, it is
logical to expect a sort of link between the entropy generated by a
quench, and the development of chaos in the Dicke model.

\begin{figure}
\begin{center}
\begin{tabular}{c}
\includegraphics[height=0.8\linewidth,angle=-90]{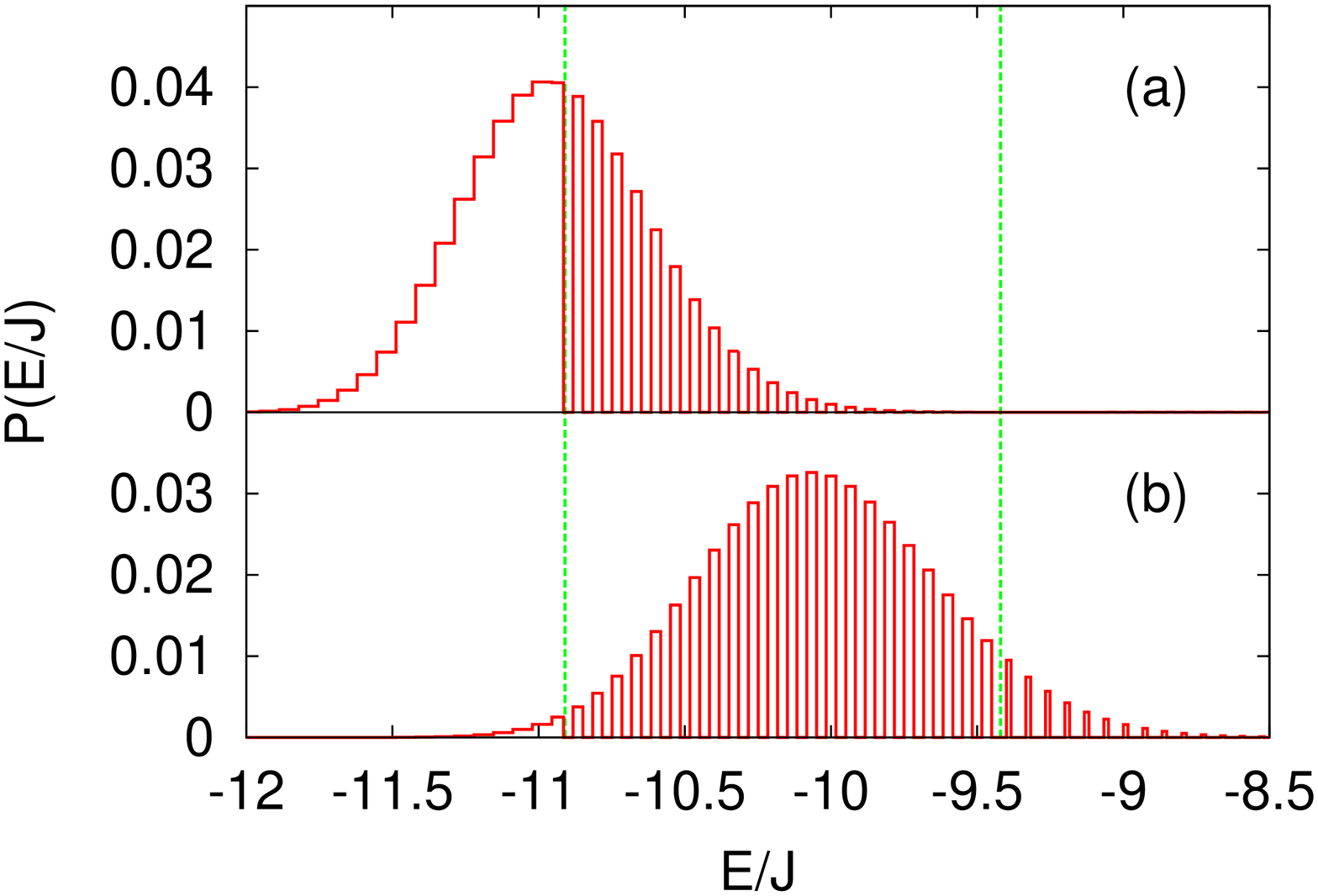} \\
\includegraphics[height=0.8\linewidth,angle=-90]{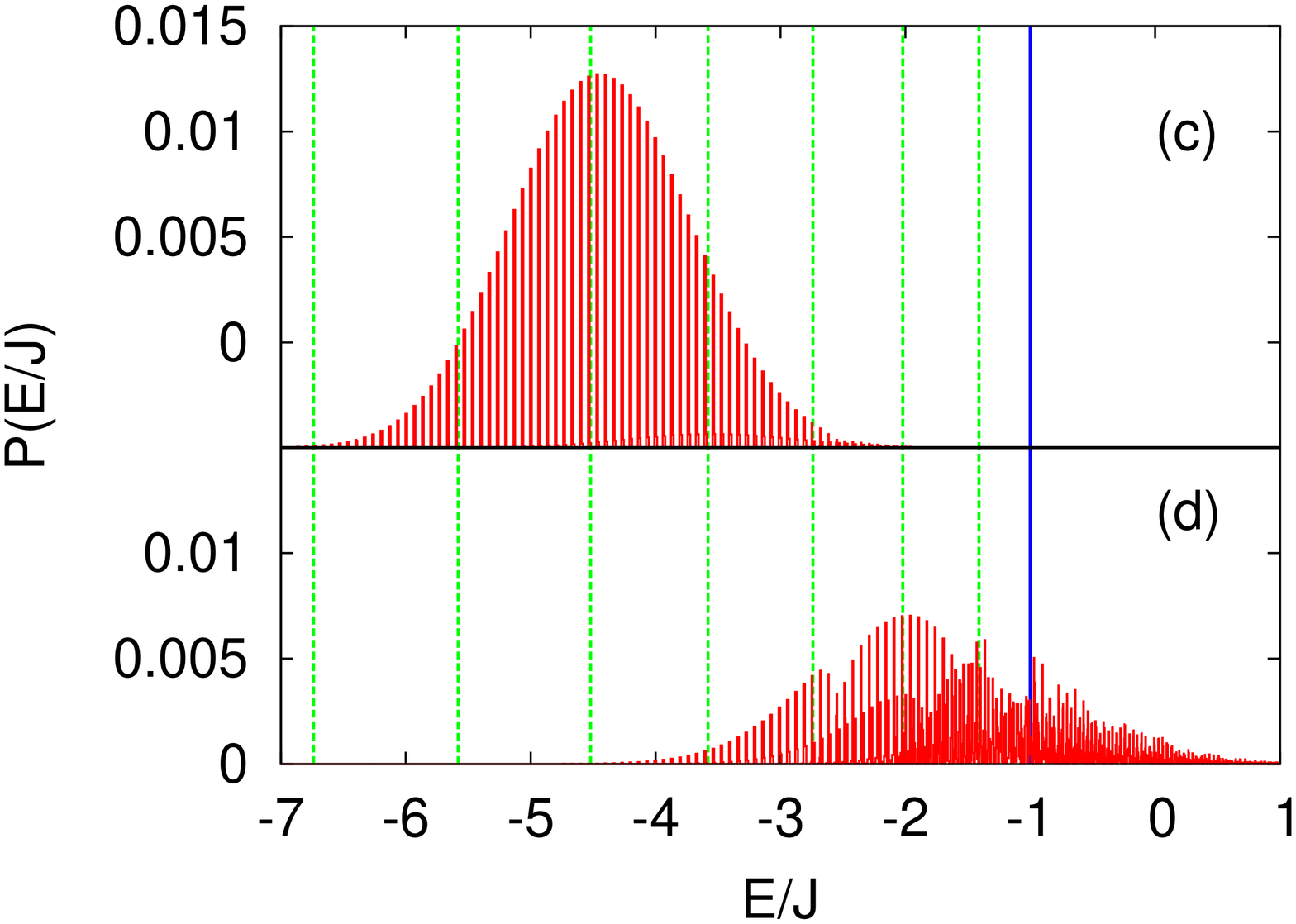} \\
\end{tabular}
\end{center}
\caption{Energy distributions $P(E_n)$ for a system with $N=30$ and
  $\lambda_f=2.5$, resulting from different quenches: panel (a),
  $\lambda_i=3.4$; panel (b), $\lambda_i=3.6$; panel (c),
  $\lambda_i=4.5$; panel (d), $\lambda_i=4.8$. Vertical dashed lines
  (green online) show the energies at which the different bands of
  Fig. \ref{fig:peres} start. Vertical solid line (blue online), shows
  the critical energy of the ESQPT, $E/J=-1$.}
\label{fig:distribuciones}
\end{figure}

With the aim of going deep into this possible link, we study the
dynamical consequences of chaos in the Dicke model. We first analyze
how the shape of the final energy distribution is affected by the
onset of chaos. In Fig. \ref{fig:distribuciones} we plot the final
energy distribution for a system with $N=30$ and $\lambda_f=2.5$,
after different quenches (see caption for details). Panels (a) and (b)
are representative of small quenches, keeping the final energy
distribution in the low-lying part of the spectrum, that is, in the
non-chaotic ordered region (see Figs. \ref{fig:peres} and
\ref{fig:chaos}). Together with the distribution, we plot the energies
at which the different bands displayed in Fig. \ref{fig:peres} start,
as vertical dashed lines (green online); in panels (a) and (b) we show
the energies corresponding to the second, $E_2/J \sim -10.9$, and the
third $E_3/J \sim -9.4$, bands. We can see that the appearance of
these bands entails qualitative changes in the energy
distributions. In the region in which there exists only one band
(between $E/J \sim -12.5$ and $E/J \sim -10.9$), all the eigenstates
become populated after the quench. On the contrary, both panels (a)
and (b) show that only one every two eigenstates become populated in
the region with two bands (between $E/J \sim -10.9$ and $E/J \sim
-9.4$). Finally, in panel (b) we can
also see that only one every three eigenstates become populated in the
region with three bands (above $E/J \sim -9.4$). This suggests that a
sort of approximated conservation rule holds: as our initial state is
the ground state of the same system with $\lambda_i > \lambda_f$, only
the eigenstates belonging to the first band are populated as a
consequence of the quench; transitions between different bands are
approximately forbidden by this conservation law.

Results for large quenches, shown in panels (c) and (d), are
different. The structure of the final eigenstate distribution is still
regular in panel (c), though the approximated conservation rule
suggested above is not so clearly seen. In fact, very small peaks
appear around $E/J \sim -4.5$, giving rise to a secondary structure
suggesting that the approximated conservation rule starts to
disappear. In panel (d), the distribution becomes erratic. In this
case, the quench leads the system to the chaotic region. We can see
that the complexity of the final distribution increases with energy,
becoming random for energies above $E_c$. The secondary structure
foreseen in panel (c) also appears around $E/J \sim -4$, and becomes
much more relevant around $E/J \sim -3$. After $E/J \sim -1.5$ we
cannot distinguish between the main and the secondary structures, and
the total distribution becomes totally erratic for energies above
$E_c$. These results suggest that the approximated conservation rules,
responsible of the regular behavior shown in panels (a) and(b), are
destroyed by chaos. When the quench leads the system to the fully
chaotic region, non-adiabatic transitions between all the energy
levels are possible.

\begin{figure}
\begin{center}
\begin{tabular}{c}
\includegraphics[height=0.8\linewidth,angle=-90]{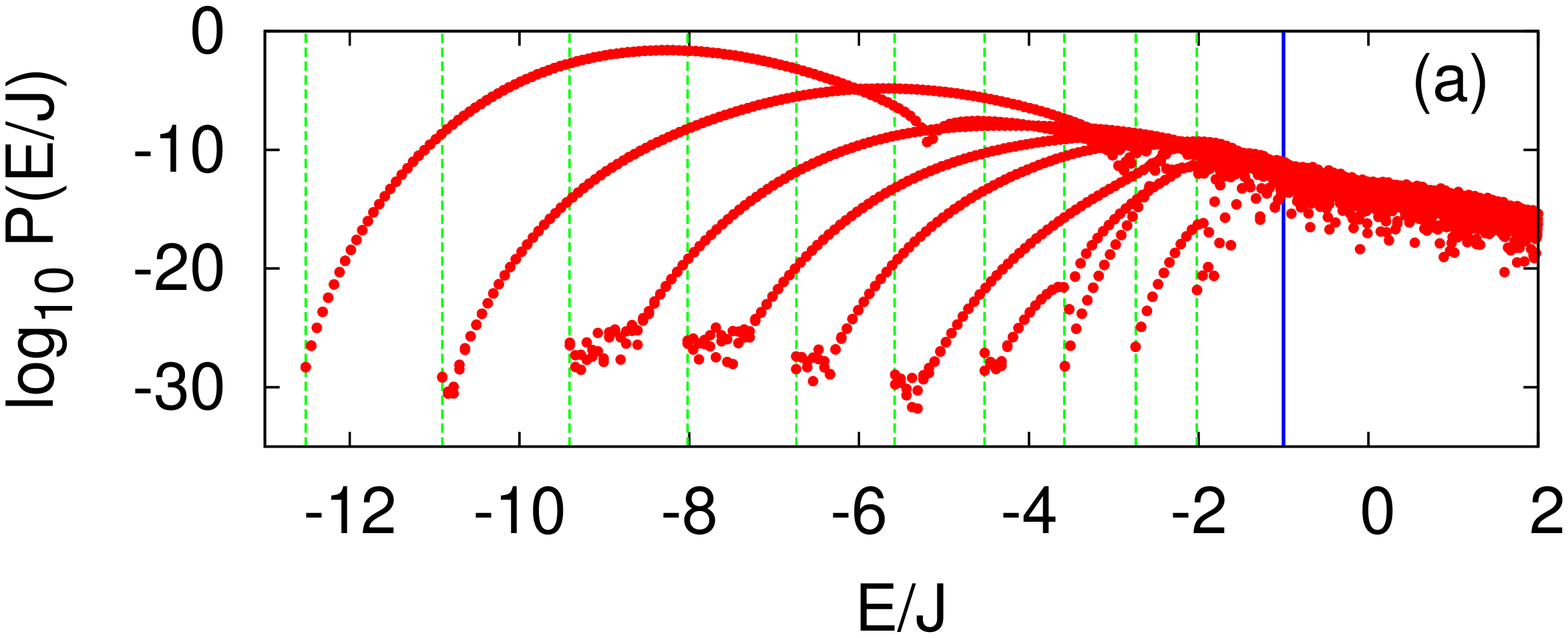} \\
\includegraphics[height=0.8\linewidth,angle=-90]{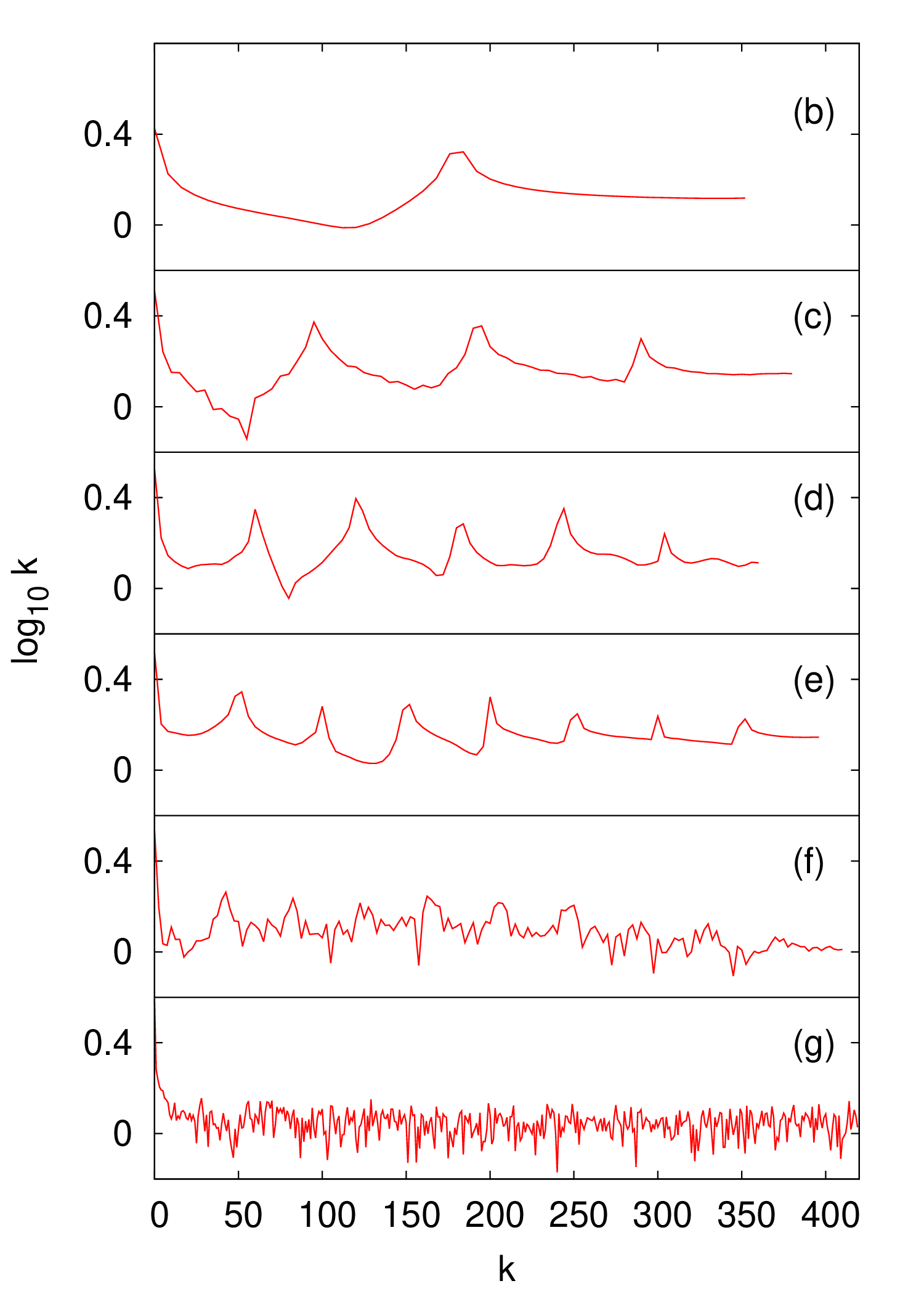}
\end{tabular}
\end{center}
\caption{Panel (a), logarithm of the final probability distribution,
  $P(E_n/J)$ for $\lambda_f=2.5$ and $N=30$. Panels (b)-(g), square
  modulus of the Fourier Transform of the following series: panel (b),
  $x^{(2)}_n$, corresponding to eigenstates in the region with two
  bands, $-10.9 \lesssim E/J \lesssim -9.4$; panel (c), $x^{(4)}$,
  corresponding to eigenstates in the region with four bands, $-8.0
  \lesssim E/J \lesssim -6.7$; panel (d), $x^{(6)}$, corresponding
  to eigenstates in the region with six bands, $-5.6 \lesssim E/J
  \lesssim -4.5$; panel (e), $x^{(8)}$, corresponding to eigenstates
  in the region with eight bands, $-3.6 \lesssim E/J \lesssim -2.7$;
  panel (f), $x^{(10)}$, corresponding to eigenstates in the region
  with ten or more bands, $-2.0 \lesssim E/J \lesssim -1.0$; panel
  (g), $x^{(11)}$, corresponding to eigenstates above the critical
  energy of the ESQPT, $-1.0 \lesssim E/J \lesssim 1.0$.}
\label{fig:fourier}
\end{figure}

To obtain a more quantitative analysis of the transition from a
regular pattern to a random behavior in the eigenstate distribution
after the quench, we proceed as follows. First, we select an
intermediate quench, $\lambda_i=4.0$, with the aim of spreading the
wavefunction over both regular and chaotic eigenstates. Second, we
calculate the logarithm of the corresponding final energy distribution
at $\lambda_f=2.5$, obtaining a series $x_n = \log_{10} P(E_n/J)$. The
logarithmic scale makes possible to study patterns and periodicities
in the final energy distribution even for very low occupation
probabilities. The result is shown in panel (a) of
Fig. \ref{fig:fourier}, together with the energies at which the
different bands start (as vertical dashed lines, green online), and
the critical energy of the ESQPT (as a vertical solid line, blue
online). We see a clearly regular pattern for the low-lying
eigenstates, as expected. But it is also worth to note that the
previously discussed conservation rules are only approximated; all the
bands are significantly populated \cite{nota2}, though only the
population of the eigenstates in the first band ir relatively large,
within a window around the expected final energy. Furthermore, we can
see the first signatures of an irregular behavior between $E/J=-4$ and
$E/J=-3$, though the plot becomes clearly noisy only above $E/J \sim
-1.5$. From this plot, we get a quantitative analysis of regularities
and patterns in the dynamics after the quench in the following way. We
divide the series $x_n \equiv \log_{10} P(E_n/J)$ in different
sequences, each one corresponding to a different region in panel (a)
of Fig.{\ref{fig:fourier}: $x^{(1)}_n$ is the logarithm of the
  probability $P(E_n)$ of the eigenstates in the region with just one
  band, $-12.5 \lesssim E/J \lesssim -10.9$; $x^{(2)}_n$, the same for
  the eigenstates in the region with two bands, $-10.9 \lesssim E/J
  \lesssim -9.4$, and so on. Finally, we obtain the square modulus of
  the Fourier Transform of the sequences $x^{(i)}_n$. Results are also
  plotted in Fig. \ref{fig:fourier}, for $x^{(2)}_n$, $x^{(4)}_n$,
  $x^{(6)}_n$, $x^{(8)}_n$, $x^{(10)}_n$, and $x^{(11)}_n$ (see
  caption for details). In panels (b)-(e) we see that the number of
  peaks in the Fourier Transform of $x^{(i)}_n$ coincides with the
  number of bands in the corresponding region of the spectrum. This
  fact is a consequence of the regular structure shown in panel (a) of
  the same figure. $x^{(2)}_n$ oscillates between the first (upper set
  of points) and the second (lower set of points) bands, and
  therefore, the corresponding Fourier Transform shows a peak in $k=0$
  and a second peak related to the frequency of these oscillations,
  which is more or less the same for the complete
  subsequence. Something similar happens with $x^{(4)}_n$. In this
  case, the signal oscillates between four different points, entailing
  three different frequencies, and the peak at $k=0$. This scenario is
  broken in the last two panels. Panel (f) lays in the region
  transiting from integrability to fully developed chaos, $-2.0
  \lesssim E/J \lesssim -1.0$. Despite a number of peaks can be still
  distinguished, the signal is noisy. Finally, panel (g) show the
  results for levels above the critical energy of the ESQPT, $-1.0
  \lesssim E/J \lesssim 1.0$. In this case, we find no traces of
  peaks; the signal is compatible with a white noise, that is, with an
  uncorrelated sequence. These facts entail that the probability
  distribution $P(E_n)$ transits from a periodic structure, coming
  from the approximated conservation rules holding in the low-lying
  region of the spectrum, to a totally uncorrelated one, occurring in
  the fully chaotic region.

\begin{figure}
\begin{center}
\includegraphics[height=0.8\linewidth,angle=-90]{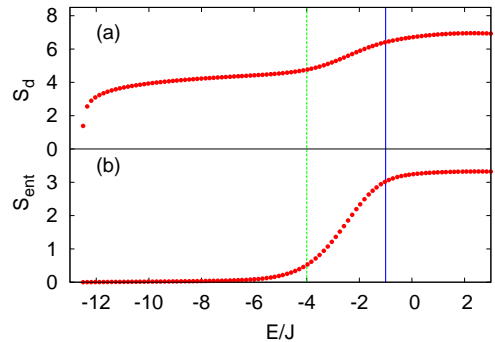}
\end{center}
\caption{Diagonal entropy $S_d$ (panel a) and entanglement entropy
  $S_{\text{ent}}$ (panel b), for a system with $N=30$ atoms and
  different values for the final coupling constants, $\lambda_f$, as a
  function of the final energy $E/J$. }
\label{fig:entro_e}
\end{figure}

All these results show that dynamics is strongly affected by the
development of chaos; the final energy distribution, which is directly
correlated with the diagonal entropy $S_d$, dramatically changes from
the almost integrable to the fully chaotic regimes. The effects in the
entropy generated by a quench are studied in
Fig. \ref{fig:entro_e}. There, we show the diagonal entropy (panel a),
and the entanglement entropy (panel b), for $N=30$ and
$\lambda_f=2.5$, as a function of the final energy $E/J$. We also show
the energies at which chaos starts, $E/J \sim -4$, as a vertical
dashed line (green online), and the critical of the ESQPT, $E/J = -1$,
as a vertical solid line (blue online). From the results, we can infer
a neat correlation between the onset of chaos and the increase of both
$S_d$ and $S_{\text{ent}}$. In particular, this correlation is
remarkable for the entanglement entropy $S_{\text{ent}}$; we can see
impressive similarities between Fig. \ref{fig:chaos} and panel (b) of
Fig. \ref{fig:entro_e}. Hence, from the comparison of these two
figures we can conclude that: {\em i)} atoms and photons remain
unentangled for quenches keeping the system in the almost-integrable
region, characterized by a number of approximated conservation rules
that prevent the majority of the non-adiabatic transitions between
energy levels; {\em ii)} the entanglement fastly increases in the
region transiting from integrability to chaos; and {\em iii)} atoms
and photons are almost yet maximally entangled after the system
crosses the ESQPT and enters in the fully chaotic regime. It is worth
to mention that these facts reinforce the link between entanglement
and thermalization \cite{Khlebnikov:14, Kaufman:16}: $S_{\text{ent}}$
starts to grow when the ordered pattern in $\left<a^{\dagger} a
\right>$ starts to disappear, that is, when the conditions for the ETH
start to hold. We can expect that quenches leading the system to the
low-lying region of the spectrum do not result in a thermal
equilibrium state; approximated conservation rules prevent ETH, and
the atomic and photonic parts of the system remains too uncorrelated
for the emergence of canonical states \cite{Khlebnikov:14}. On the contrary,
quenches leading the system to the chaotic region, in particular to
final energy values above the ESQPT, are expected to entail a thermal
behavior; no regularities in the final energy distribution are found,
and the entanglement between the atomic and the photonic parts of the
system is very large.

The diagonal entropy $S_d$ is also linked to the onset of chaos, but
not so strongly.  Its fastly-growing region also coincides with the
transition from integrability to chaos. We can interpret this result
as a consequence of the breaking of the approximated conservation
rules holding in the almost-integrable region. When this transiting
region is reached, a lot of non-adiabatic transitions become possible,
spreading the final energy distribution over a much larger number of
eigenstates, and therefore largely increasing the diagonal
entropy. However, $S_d$ is also significantly larger than zero for
quenches leading the system to the almost-integrable region. This fact
entails that such quenches are irreversible, though they neither
correlate atoms and photons, nor lead the system to a thermal
behavior.

\subsection{ESQPT}

Results in previous sections suggest that the emergence of chaos plays
a prominent role in the entanglement between the atomic and the
photonic part of the Dicke model. Notwithstanding, the ESQPT seems to
be important too. Results shown in Figs. \ref{fig:chaos},
\ref{fig:fourier} and \ref{fig:entro_e} suggest that the system
becomes fully chaotic after crossing the critical energy of the ESQPT,
atoms and photons become maximally entangled roughly at the same
point, and also the fastly-growing region of the diagonal entropy
approximately ends at the same value. Since the results for
$S_{\text{ent}}$ are more clear than for $S_d$ (the entanglement
entropy grows from zero to almost its maximum possible value in a very
narrow energy window), we explore now the possible links between this
magnitude and the ESQPT.

\begin{figure}
\begin{center}
\includegraphics[height=0.8\linewidth,angle=-90]{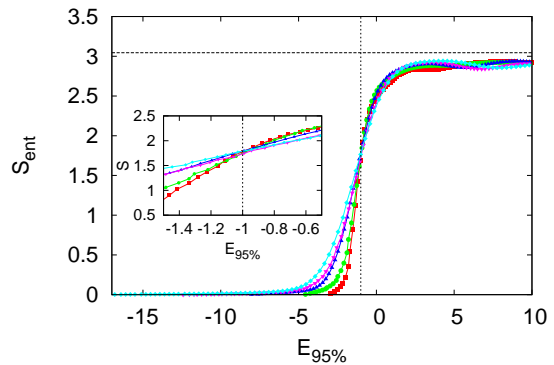} 
\end{center}
\caption{Entanglement entropy $S_{\text{ent}}$ for $N=20$ and different values of $\lambda_f$, as a function of $E_{95\%}$. (See main text
  for details). Squares (red online) corresponds to $\lambda_f=1.2$;
  circles (green online), to $\lambda_f = 1.5$; lower triangles (blue
  online), to $\lambda_f=2.0$; lower triangles (magenta online), to
  $\lambda_f=2.5$, and diamonds (cyan online), to
  $\lambda_f=3.0$. Vertical dotted line shows the critical energy of
  the ESQPT, $E_c/J=-1$; horizontal dashed line, the maximum possible
  value for the von Neumann entropy, $S=\log 21$. The inset shows a
  zoom of the same results around $E/J=-1$.}
\label{fig:cruces}
\end{figure}

Results shown in Fig. \ref{fig:entro_e} are not conclusive regarding
the possible link between entanglement and the ESQPT. The development
of chaos occurs within a quite wide energy window; hence, the fact
that the final energy distribution is also wide does not constitute a
problem. On the contrary, the ESQPT is a critical phenomenon, taking
place at a precise value of the energy, $E_c/J=-1$; thus, we can
expect that its dynamical consequences become blurred if the final
energy distribution is wide. To avoid this problem we recalculate the
consequences of the same quenches, though for a smaller system with
$N=20$ (to have more numerical points), and we study the entanglement
entropy $S_{\text{ent}}$ in terms of $E_{95 \%}$.  This is the energy
value which fulfills $F(E \leq E_{95 \%})=0.95$, being $F(E)$ the
cumulative probability distribution for the energy in a
measurement. In other words, we plot the results versus the energy
below which the $95 \%$ of the energy distribution lays. This value
has been selected to identify the point at which the system {\it
  touches} the critical point, that is, the point at which the normal
phase starts to be significantly populated \cite{nota3}. In Fig.
\ref{fig:cruces} we plot the corresponding results for different
values of the final coupling constant, $\lambda_f$ (see caption of
Fig. \ref{fig:cruces} for details). In the inset, we plot a zoom of
the same curves, centered at the critical energy $E_{95 \%}/J=-1$. The
main conclusion is that all the five curves cross at the critical
point $E_{95 \%}/J \sim -1$, suggesting that this point plays a
significant role in the dynamics of the entanglement
entropy. Notwithstanding, this result is not as conclusive as the
obtained regarding the onset of chaos. More work is needed to separate
the consequences of chaos and the ESQPT.

\subsection{Entropy growth}

As it is discussed in the introduction, one of the main differences
between diagonal $S_d$ and entanglement $S_{\text{ent}}$ entropies is
that the first one changes immediately after the quench, while the
second one has a non-trivial time dependence, which can be related to
the Second Law \cite{Esposito:10}. We explore here this time
dependence.

\begin{figure}
\begin{center}
\begin{tabular}{cc}
  \includegraphics[width=0.25\linewidth,height=0.5\linewidth,angle=-90]{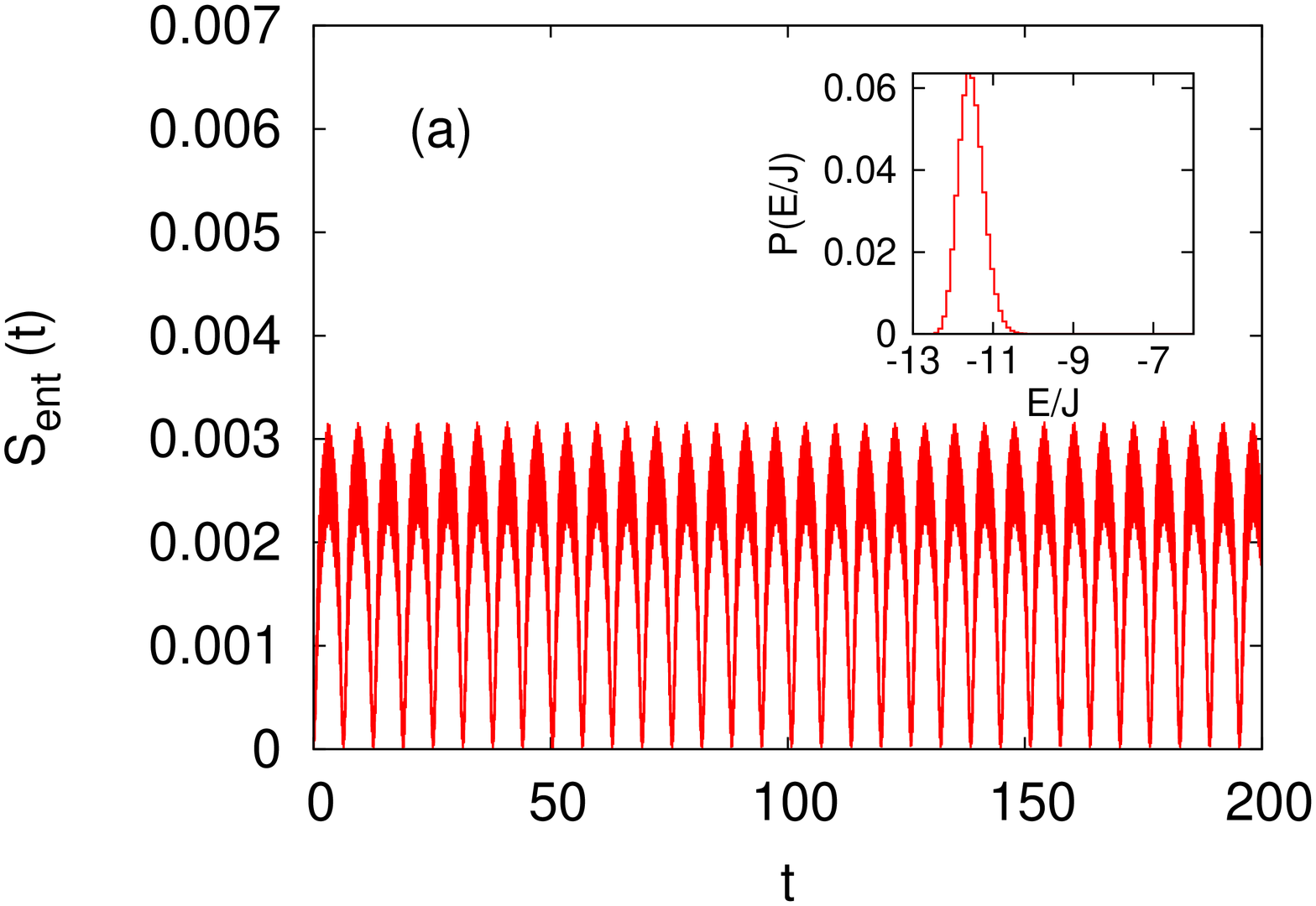} &
  \includegraphics[width=0.25\linewidth,height=0.5\linewidth,angle=-90]{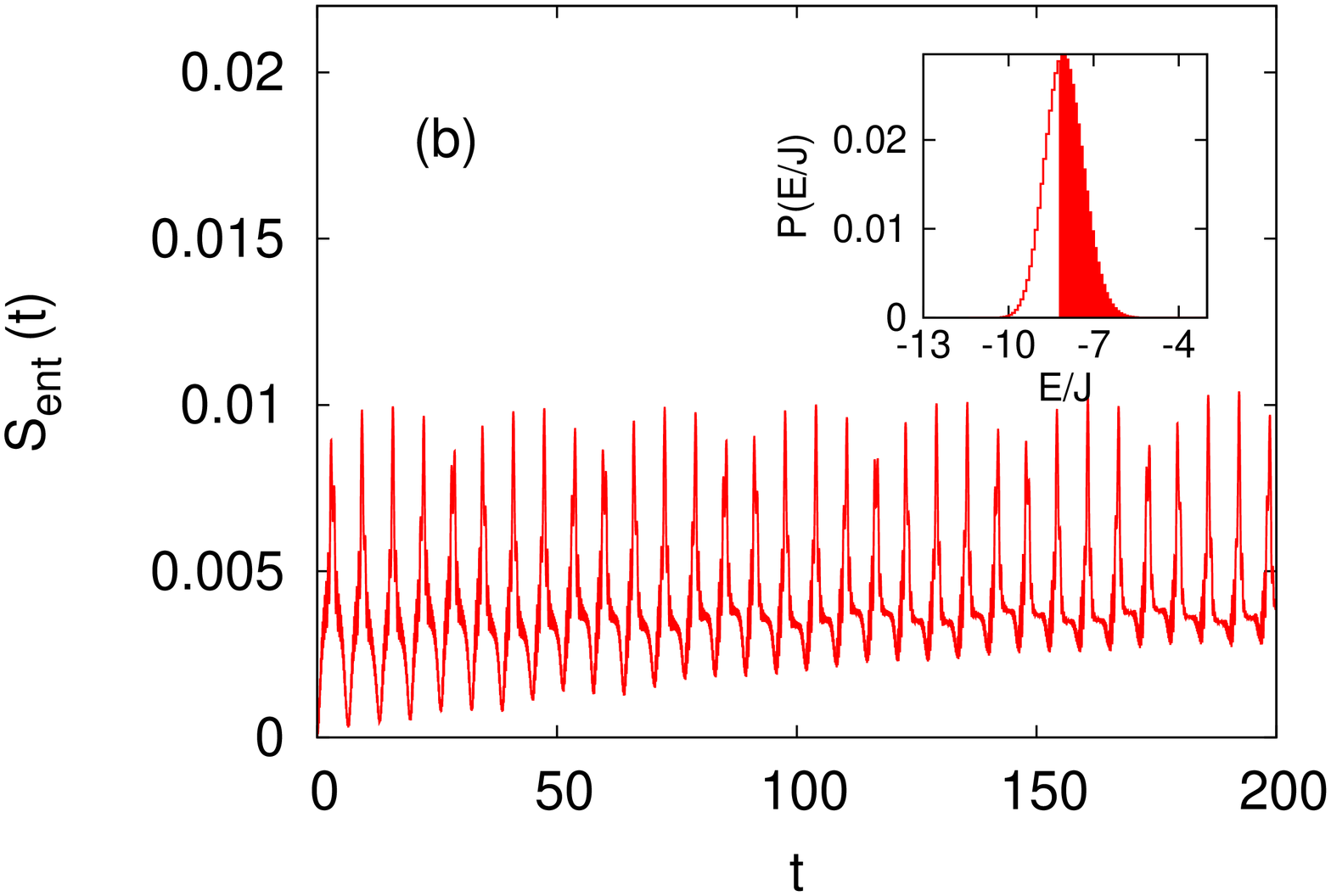} \\
  \includegraphics[width=0.25\linewidth,height=0.5\linewidth,angle=-90]{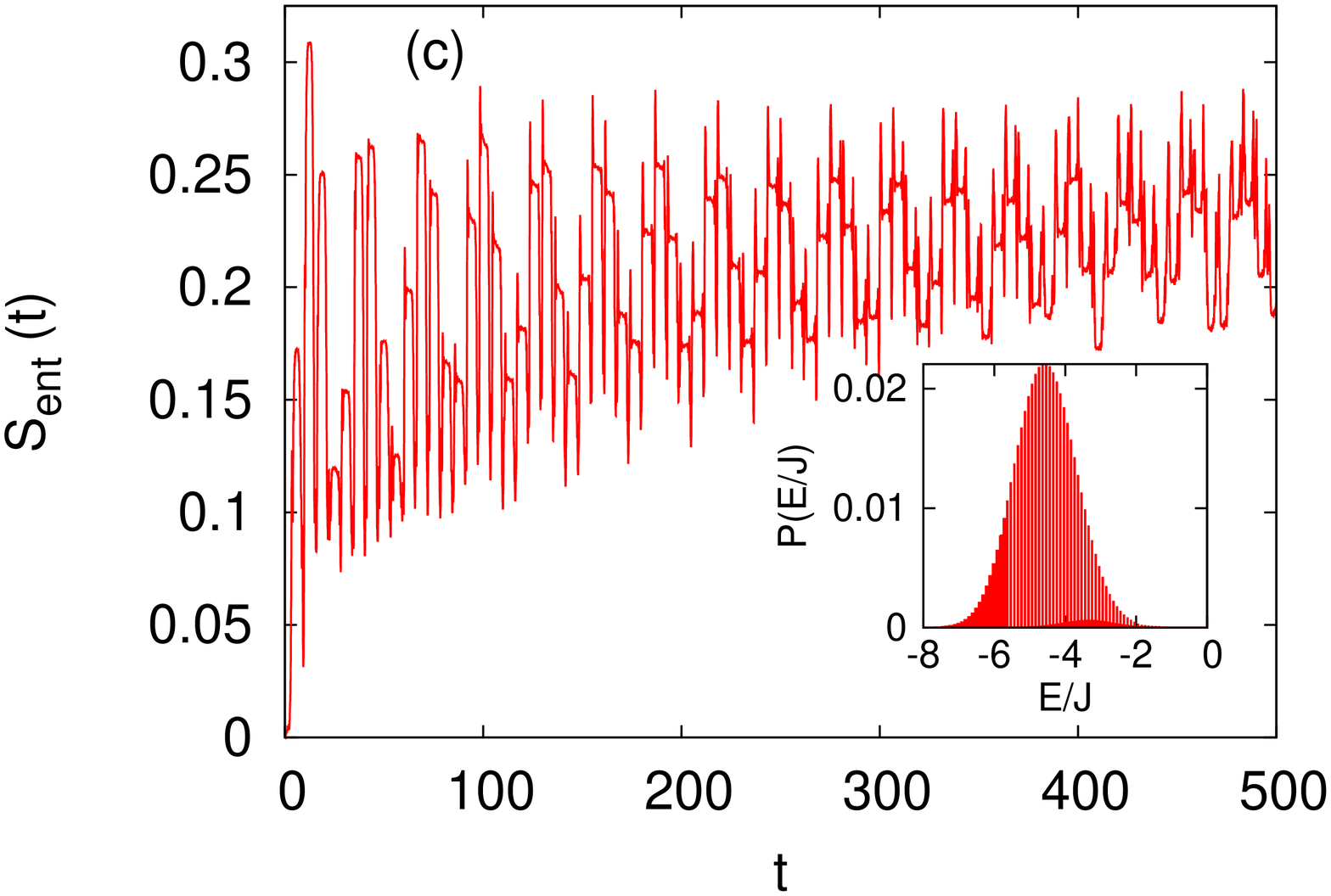} &
  \includegraphics[width=0.25\linewidth,height=0.5\linewidth,angle=-90]{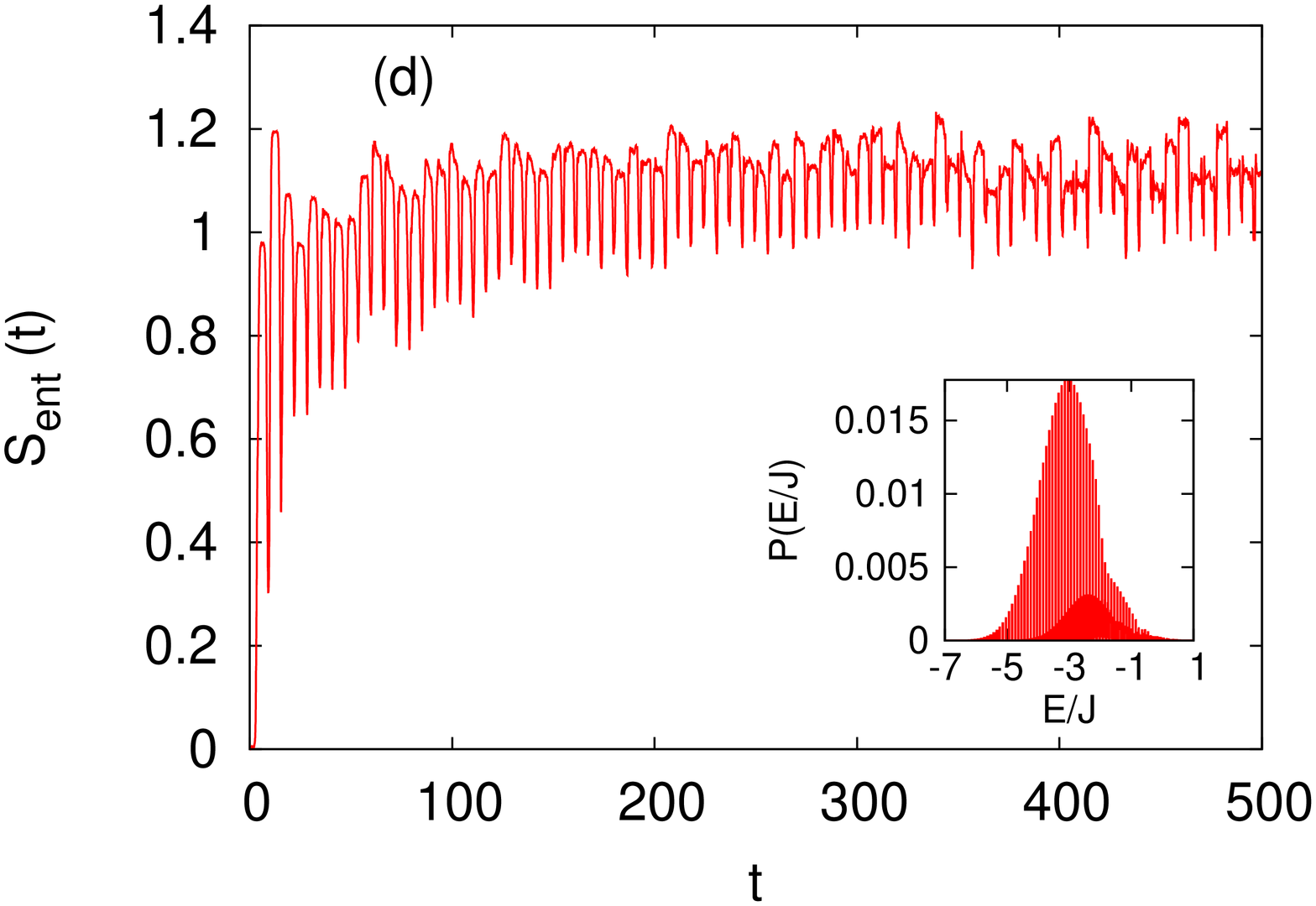} \\
  \includegraphics[width=0.25\linewidth,height=0.5\linewidth,angle=-90]{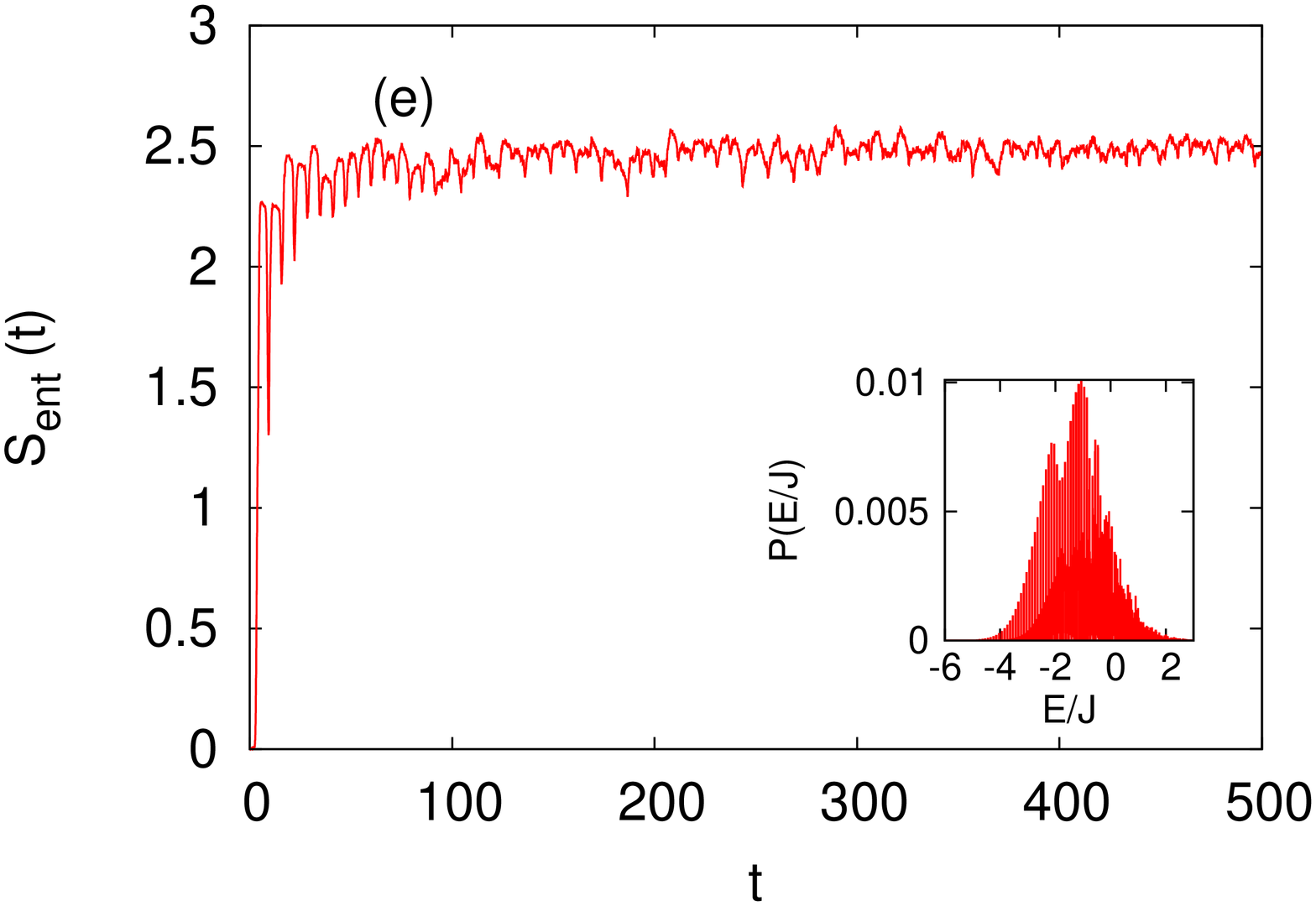} &
  \includegraphics[width=0.25\linewidth,height=0.5\linewidth,angle=-90]{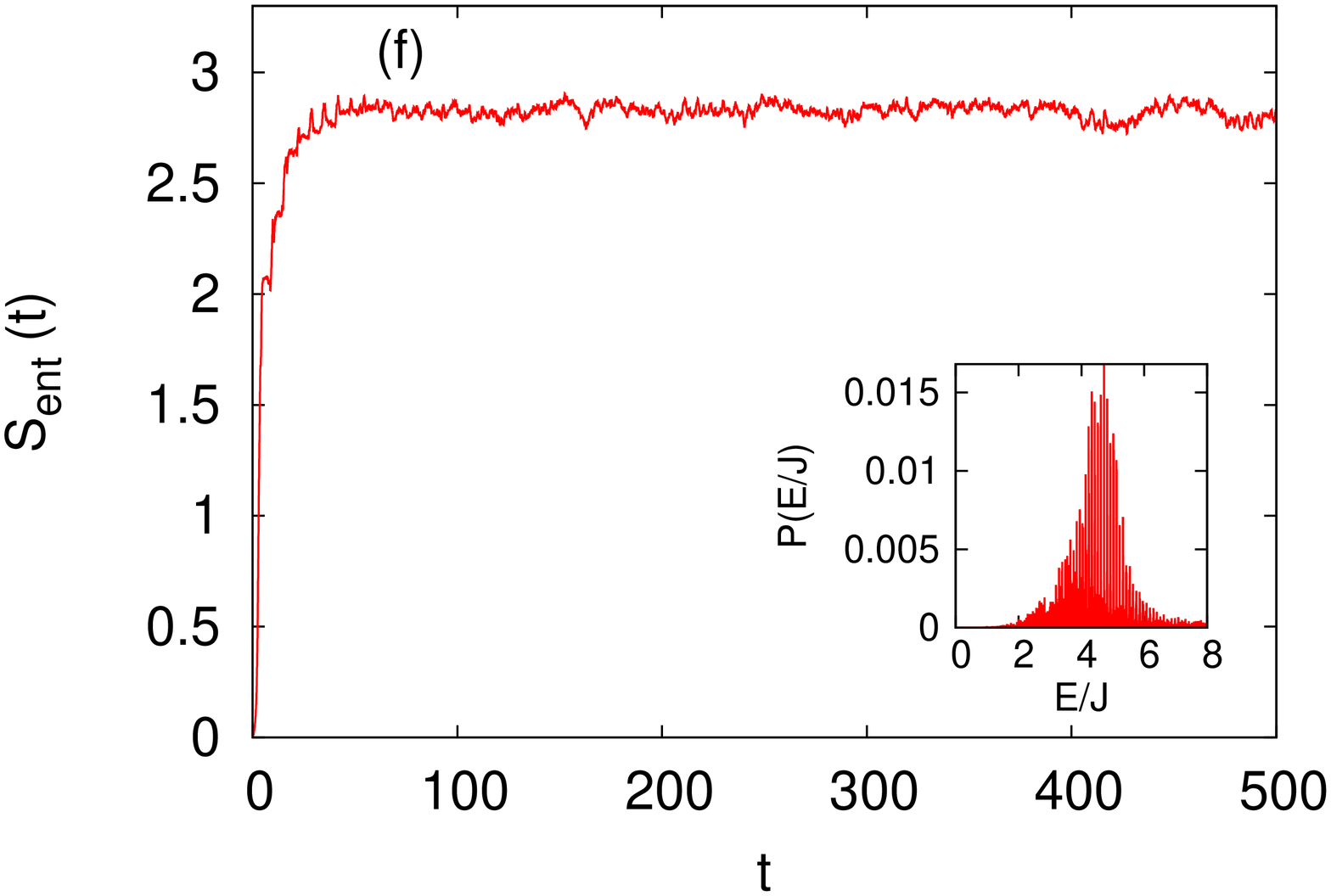} \\
\end{tabular}
\end{center}
\caption{Entropy production for a system with $N=20$ atoms and
  $\lambda_f = 2.5$. Panel (a) corresponds to $\lambda_i=3.0$; panel
  (b), to $4.0$; panel (c), to $4.7$; panel (d), to $4.9$; panel (e),
  to $5.5$, and panel (f), to $7.0$.  In the main part of the panels,
  red lines show $S_{\text{ent}}$. In the insets, it is shown the
  final energy distribution.}
\label{fig:time}
\end{figure}

In Fig. \ref{fig:time} we plot how the entanglement entropy
$S_{\text{ent}}$ grows after the quench. All the results are
numerically obtained with $N=20$ atoms and $\lambda_f = 2.5$. For the
sake of clarity, we have selected a set of representative cases which
give rise to final states with the following properties: panel (a),
with $\lambda_i=3.0$, show a case in which only the first band of
levels is populated; panel (b), with $\lambda_i=4.0$, is clearly below
the onset of chaos, though more of one band is populated; in panel
(c), with $\lambda_i=4.7$, levels within the transition from
integrability to chaos are yet populated, though almost all the energy
distribution is below the critical energy of the ESQPT, $E/J=-1$;
panel (d), with $\lambda_i=4.9$, shows a case in which the final
energy distribution {\em touches} the critical energy, and almost all
is within the transiting region; in panel (e), with $\lambda_i=5.5$,
the energy distribution is roughly centered at the critical energy
$E/J=-1$, and panel (f), with $\lambda_i=7.0$, shows a case in which
almost all the energy distribution is in the normal phase. In panel
(a), $S_{\text{ent}}$ is periodic in time, oscillating from zero to a
very small value. Panel (b) is very similar, though we can foresee a
very small entropy growth. As both cases correspond to a final state
in the low-lying regular region of the spectrum, we can conclude that
almost no entanglement is produced under these circumstances, and that
the time dependence of the corresponding entropy is not related with
the irreversibility produced by the quench. As we can see in both
panels, the final energy distribution is wide, implying that a number
of non-adiabatic transitions have occurred. Results plotted in panel
(a) of Fig. \ref{fig:entro_e} show that $S_d$ is significantly larger
than zero under these circumstances. So, these results make evident
the different roles played by entanglement and diagonal
entropies. Though the atomic part of the system remains almost pure,
entailing no entanglement entropy, the non-zero value of the diagonal
entropy indicates that we cannot revert the process. The system enters
in the region transiting from integrability to chaos in panel (c). In
this case, we can see that the main trend of $S_{\text{ent}}$ is
clearly increasing; in other words, the purity of the atomic system
decreases in time. The qualitative behavior of panel (d) is very
similar, though the final equilibrium value for $S_{\text{ent}}$ is
clearly larger, because the final energy distribution lays in the
transiting region. There is an abrupt increase of
such equilibrium values between panels (d) and (e), that is, when the
final energy distribution crosses the critical energy, and enters in
the totally chaotic region. Note that the curve in panel (e) reach its
final equilibrium value clearly earlier than the curve in panel (d),
and that the relative weight of its fluctuations is much
smaller. Finally, when almost all the energy distribution lays in the
normal phase and in the fully chaotic region, panel (f), fluctuations
are even smaller, but the main trend of $S_{\text{ent}}$ is very
similar to the previous case.

From these results we conclude that $S_{\text{ent}}$ shows a main
trend qualitatively, but not quantitatively, compatible with the
Second Law. Besides fluctuations, which are large due to the small
sizes accessible to current computational capabilities,
$S_{\text{ent}}$ grows in time, in a similar way than in recent
theoretical calculations for other systems \cite{Esposito:10} and
experiments \cite{Kaufman:16}. Starting from a separable state, and
for quenches large enough, the atomic part of the system changes from
a pure to a mixed state. However, this growth is only significant
above the onset of chaos. When the state of the system remains within
the low-lying regular part of the spectrum, the resulting entanglement
entropy is very low, or even periodic in time, despite the final
energy distribution has a finite width. In other words, below the
onset of chaos the non-adiabatic transitions resulting from the quench
do not correlate the atomic and the photonic parts of the system. So,
the entanglement entropy is not enough to characterize all the
irreversible processes taking place in a globally isolated quantum
system.

\section{Conclusions}
\label{conclusions}

In this paper we have studied the entropy generated by a quench in an
isolated quantum many-body system, the Dicke model, describing the
interaction between a set of two-levels atoms with a monocromatic
radiation field. We have dealt with different magnitudes: the diagonal
entropy, $S_d$, which can be understood as the von Neumann entropy of
the final equilibrium state, $\rho_{\text{eq}}$; and the entanglement
entropy, $S_{\text{ent}}$, which is the von Neumann entropy of the
atomic part of the system, obtained by tracing out the photonic
degrees of freedom. We have computed numerically the exact
time-evolution of the complete system, and we have derived both
entropies from these exact results. We have obtained the following main
conclusions.

First, there is no direct link between the energy dissipated by the
quench and the increase of both kind of entropies. We have numerically
and analytically shown that the dissipated heat $Q$ just depends on
the size of the quench $\Delta \lambda = \lambda_i - \lambda_f$,
independently of the final value of the coupling constant,
$\lambda_f$. On the contrary, both kind of entropies largely depend on
this parameter. We have shown that the same amount of dissipated
energy $Q$ can entail an entropy production either very large (for
small values of $\lambda_f$), or very small (for large values of
$\lambda_f$). Hence, despite the dissipated heat just depends on the
size of the quench $\Delta \lambda$, the entropy production
dramatically depends on the structure of the spectrum of the final
Hamiltonian.

Second, the entropy production is largely increased by chaos. Small
quenches, keeping the system in an almost-integrable regime, entail a
small value for the diagonal entropy $S_d$, and an almost-zero value
for the entanglement entropy $S_{\text{ent}}$. On the contrary, large
quenches, leading the system to a fully chaotic regime, entail very
large values for both entropies. In particular, $S_{\text{ent}}$
changes from almost-zero to almost its maximum possible value in the
same regime in which the final Hamiltonian changes from
almost-integrable to fully chaotic. It is worth to mention that this
fact is closely related to the transition from non-thermal to thermal
behavior. In the almost-integrable regime the conditions of the
Eigenstate Thermalization Hypothesis (ETH), the mechanism responsible
for thermalization in isolated quantum systems, does not hold, so we
cannot expect that the system behaves according to standard
thermodynamics.  On the contrary, ETH holds when the dynamics becomes
fully chaotic, so we can expect that the system {\em forgets} all the
details about its initial condition in this regime, giving rise to a
thermal final equilibrium state. We have shown that the transition in
the entanglement entropy $S_{\text{ent}}$ is highly correlated with
the transition from non-ETH to ETH scenarios. This result is
compatible with recent experiments, showing that the entanglement
between different parts of a (small) quantum system is behind
thermalization \cite{Kaufman:16}. Regarding the diagonal entropy, we
have shown that a similar transition also takes place, though the
change is not so abrupt as for $S_{\text{ent}}$. We conclude
that the process becomes more irreversible when the system enters in
the fully chaotic regime, because in the almost-integrable regime
there are a number of approximated conservation rules, limiting the
amount of non-adiabatic transitions between energy levels.

Third, there is a subtle link between the entanglement entropy,
$S_{\text{ent}}$, and the presence of an excited-state quantum phase
transition (ESQPT). We have shown that the critical energy of the
ESQPT, $E_c$, is also a singular point regarding the entanglement
entropy production. However, this link is clearly weaker than the
previous one, and requires further research.

Finally, we have shown that the time dependence of $S_{\text{ent}}$
qualitatively agrees with a thermodynamical entropy supporting
the Second Law, but this agreement is not quantitatively correct. When
the final state lays in the almost-integrable regime, atoms and
photons remain almost unentangled, though the process is irreversible,
a certain amount of energy is dissipated, and the diagonal entropy
$S_d$ is significantly larger than zero. It is worth to remark that
this occurs in a regime in which thermalization is not expected, that
is, in a regime in which standard thermodynamics does not work.

Summarizing, we have found a clear link between irreversibility and
chaos in the Dicke model. Also, we have found some traces pointing a
weaker link between entropy production and crossing the critical point
of an ESQPT. In any case, we conclude that the dissipated energy is
directly correlated neither with the diagonal nor with the
entanglement entropy generated by a quench. Hence, the recently
proposed link between the growth of entanglement between a system and
its environment and the Second Law \cite{Esposito:10} has to be worked
with some caution.

\section{Aknowledgements}

This work has been partially supported by the Spanish Mineco Grant
No. FIS2012-35316. C. M. L. is supported by the Spanish MECD Program
'Becas de Colaboraci\'on'. A. R. thanks R. Puebla for enlightening
discussions. C. M. L. thanks A.Alsina and D.Mendez-Gonzalez for useful
advices that clarified the meaning of several topics.


\section{References}

\begin{thebibliography}{99}

\bibitem{Bloch:08} I. Bloch, J. Dalibard, and W. Zwerger, Rev. Mod. Phys. {\bf 80}, 885 (2008).

\bibitem{Polkovnikov:11b} A. Polkovnikov, K. Sengupta, A. Silva, and
  M. Vengalatore, Rev. Mod. Phys. {\bf 83}, 863 (2011).

\bibitem{Deutsch:91} J. M. Deutsch, Phys. Rev. A {\bf 43}, 2046 (1991).

\bibitem{Srednicki:94} M. Srednicki, Phys. Rev. E {\bf 50}, 88 (1994).

\bibitem{Rigol:08} M. Rigol, V. Dunjko, and M. Olshanii, Nature {\bf
    452}, 854 (2008).

\bibitem{QuantumThermodynamics} J. Gemmer, M. Michel, G. Mahler, {\em
    Quantum Thermodynamics}, Lect. Notes Phys. {\bf 657} (Springer,
  Berling Heidelberg 2005).

\bibitem{Esposito:10} M. Esposito, K. Lindenberg, and
  C. Van den Broeck, New Jour. Phys. {\bf 12}, 013013 (2010).

\bibitem{Allahverdyan:05} A. E. Allahverdyan and Th. M. Nieuwenbuizen, Phys. Rev. E {\bf 71}, 046107 (2005).

\bibitem{Puebla:15} R. Puebla and A. Rela\~no, Phys. Rev. E {\bf 92},
  012101 (2015).

\bibitem{Polkovnikov:05} A. Polkovnikov, Phys. Rev. B {\bf 72}, R161201 (2005).

\bibitem{Zurek:05} W. H. Zurek, U. Dorner, and P. Zoeller, Phys. Rev. Lett. {\bf 95}, 105701 (2005).

\bibitem{Grandi:09} C. De Grandi and A. Polkovnikov, arXiv:0910.2216 (2009).

\bibitem{Dicke:54} R. H. Dicke, Phys. Rev. {\bf 93}, 99 (1954).

\bibitem{termo} H. B. Callen, {\it Thermodynamics and an Introduction to Thermostatistics}, (John Willey $\&$ sons, Inc. 1985).

\bibitem{Polkovnikov:11} A. Polkovnikov, Ann. Phys. (N. Y.) {\bf 326}, 486 (2011).

\bibitem{Reimann:12} P. Reimann and M. Kastner, New. J. Phys. {\bf 14}, 043020 (2012).

\bibitem{review} R. Horodecki, P. Horodecki, M. Horodecki, and
  K. Horodecki, Rev. Mod. Phys {\bf 81}, 865 (2009); L. Amico,
  R. Fazio, A. Osterloh, and V. Vedral, Rev. Mod. Phys, {\bf 80}, 517
  (2008).

\bibitem{Popescu:06} S. Popescu, A. J. Short, and A. Winter,
  Nat. Phys. {\bf 2}, 754 (2006).

\bibitem{Khlebnikov:14} S. Khlebnikov and M. Kruczenski, Phys. Rev. E {\bf 90}, 050101(R) (2014).

\bibitem{Kaufman:16} A. M. Kaufman, M. E. Tai, A. Lukin, M. Rispoli,
  R. Schittko, P. S. Preissm and M. Greiner, arXiv:1603.04409 (2016).

\bibitem{Baumann:10} K. Baumann, C. Guerlin, F. Bennecke, and
  T. Esslinger, Nature {\bf 464}, 1301 (2010).

\bibitem{Solano:14} A. Mezzacapo, U. Las Heras, J. S. Pedernales,
  L. DiCarlo, E. Solano, and L. Lamata, arXiv:1405.581 (2014).

\bibitem{Cejnar:06} P. Cejnar, M. Macek, S. Heinze, J. Jolie, and. J. Dobes, J. Phys. A {\bf 39}, L515 (2006).

\bibitem{Caprio:08} M. A. Caprio, P. Cejnar, and F. Iachello, Ann. Phys. (N. Y.) {\bf 323}, 1106 (2008).

\bibitem{Stransky:14} P. Stransky, M. Macek, and P. Cejnar, Ann. Phys. (N. Y.) {\bf 345}, 73 (2014).

\bibitem{Perez-Fernandez:11b} P. P\'erez-Fern\'andez, A. Rela\~no,
  J. M. Arias, P. Cejnar, J. Dukelsky, and J. E. Garc\'{\i}a-Ramos,
  Phys. Rev. E {\bf 83}, 046208 (2011).

\bibitem{Bastarrachea:14} M. A. Bastarrachea-Magnani, S. Lerma-Hern\'andez, and J. G. Hirsch, Phys. Rev. A {\bf 89}, 032102 (2014).

\bibitem{Emary:03} C. Emary and T. Brandes,
  Phys. Rev. Lett. 90, 044101 (2003); Phys. Rev. E 67, 066203 (2003).

\bibitem{Hepp:73} K. Hepp and E. H. Lieb, Ann. Phys. (NY) {\bf 76},
  360 (1973).

\bibitem{Wang:73} Y. K. Wang and F. T. Hioe, Phys. Rev. A {\bf 7}, 831
  (1973). 

\bibitem{Carmichael:73} H. J.  Carmichael, C. W. Gardiner, and D. F. Walls,
  Phys. Lett. A {\bf 46}, 47 (1973).

\bibitem{Lambert:04} N. Lambert, C. Emary, and T. Brandes, Phys. Rev. Lett. {\bf 92}, 073602 (2004).

\bibitem{Perez-Fernandez:11} P. P\'erez-Fern\'andez, P. Cejnar, J. M. Arias,
  J. Dukelsky, J. E. Garc\'{\i}a-Ramos, and A. Rela\~no, Phys. Rev. A
  {\bf 83}, 033802 (2011).

\bibitem{Puebla:13} R. Puebla, J. Retamosa, and
  A. Rela\~no, Phys. Rev. A {\bf 87}, 023819 (2013).

\bibitem{Puebla:13b} R. Puebla and
  A. Rela\~no, Europhys. Lett. {\bf 104}, 50007 (2013).

\bibitem{Brandes:13} T. Brandes,
  Phys. Rev. E {\bf 88}, 03213 (2013).

\bibitem{Relano:14} A. Rela\~no, J. Dukelsky, J. M. Arias, and P. P\'erez-Fern\'andez, Phys. Rev. E {\bf 90}, 042139 (2014).

\bibitem{Zhang:90} W.- M. Zhang, D. H. Feng, and R. Gilmore, Rev. Mod. Phys. {\bf 62}, 867 (1990).

\bibitem{Quan:05} H. T. Quan, P. Zhang, and C. P. Sun, Phys. Rev. E
  {\bf 72}, 056110 (2005)

\bibitem{Polkovnikov:08} A. Polkovnikov, Phys. Rev. Lett. {\bf 101}, 220402 (2008).

\bibitem{Gomez:11} J. M. G. G\'omez, K. Kar, V. K. B. Kota, R. A. Molina, A. Rela\~no, and J. Retamosa, Phys. Rep. {\bf 499}, 103 (2011).

\bibitem{Stransky:09} P. Str\'ansk\'y, P. Hru\v{s}ka, and P. Cejnar, Phys. Rev. E {\bf 79}, 066201 (2009).

\bibitem{Peres:84} A. Peres, Phys. Rev. Lett. {\bf 53}, 1711 (1984).

\bibitem{nota1} The original method by Peres is devoted to the study
  of the transition from integrability to chaos, by perturbing an
  integrable Hamiltonian $H_0$, giving rise to $H(\lambda)=H_0 +
  \lambda V$. If $\lambda=0$ there exists at least one non-trivial
  integral of motion $I$ commuting with the Hamiltonian,
  $[H,I]=0$. This entails that, in this case, the plot of $\left<E_n
  \right| I \left| E_n \right>$ versus the corresponding energy is
  regular. On the contrary, if $\lambda \neq 0$, $I$ is no longer
  an integral of motion, and hence the order in the same plot is
  destroyed. For this reason, in the original Peres Lattices the
  chosen observable has to be an integral of motion in a certain
  limit, $\lambda=0$. It is worth to note that the observable we have
  chosen in Fig. \ref{fig:peres}, the number of photons $a^{\dagger}
  a$ is an integral of motion if $\lambda=0$.

\bibitem{nota2} As calculations are performed in double precision,
  coefficents up to $10^{-15}-10^{-16}$ in the expansion of the
  wavefunction in the eigenbasis of the final Hamiltonian $\lambda_f$
  are significant, entailing probabilities up to
  $10^{-30}-10^{-32}$.

\bibitem{nota3} The choice of $E_{95 \%}$ is arbitrary; we could
  choose $E_{99 \%}$, $E_{ 90 \%}$, or any other value representing
  the tail of the distribution. We can understood that the $95 \%$ is
  somehoy equivalent to a statistical significance $p=0.05$.

\end{thebibliography}
\end{document}